\documentclass[english]{article}
\usepackage{custom_tex}

\title{\Large An Exact Sampler for Inference after Polyhedral Model Selection}
\date{August 2023}
\author{Sifan Liu\thanks{The author thanks Prof. Art Owen, Snigdha Panigrahi, and Jonathan Taylor for helpful conversations. This work was partially funded by the NSF grant DMS-2152780 and the Stanford Data Science Scholars program. Correspondence email: \texttt{sfliu@stanford.edu}}}
\newcommand{\wtQ}{\widetilde{Q}}
\newcommand{\betamle}{\hat{\boldsymbol{\beta}}^{\text{sMLE}}}
\newcommand{\bomega}{\boldsymbol{\omega}}
\newcommand{\bbeta}{\boldsymbol{\beta}}
\newcommand{\bfeta}{\boldsymbol{\eta}}
\newcommand{\btau}{\boldsymbol{\tau}}
\newcommand{\bzero}{\mathbf{0}}
\newcommand{\whM}{\widehat{M}}

\affil{Department of Statistics, Stanford University}

\begin{document}
\maketitle

\begin{abstract}
Inference after model selection presents computational challenges when dealing with intractable conditional distributions. Markov chain Monte Carlo (MCMC) is a common method for sampling from these distributions, but its slow convergence often limits its practicality. In this work, we introduce a method tailored for selective inference in cases where the selection event can be characterized by a polyhedron. The method transforms the variables constrained by a polyhedron into variables within a unit cube, allowing for efficient sampling using conventional numerical integration techniques. Compared to MCMC, the proposed sampling method is highly accurate and equipped with an error estimate. Additionally, we introduce an approach to use a single batch of samples for hypothesis testing and confidence interval construction across multiple parameters, reducing the need for repetitive sampling. Furthermore, our method facilitates fast and precise computation of the maximum likelihood estimator based on the selection-adjusted likelihood, enhancing the reliability of MLE-based inference. Numerical results demonstrate the superior performance of the proposed method compared to alternative approaches for selective inference.
\end{abstract}




\section{Introduction}

Inference after model selection must account for selection bias. One approach to correct the selection bias is to condition on the event of model selection. Through this conditioning, information utilized for selection is discarded, leaving only residual information for inference. In particular, inference is conducted based on the distribution of the data conditional on the selection event.

A preeminent model selection technique in regression is the lasso \citep{tibshirani1996regression}. This method entails minimizing the negative log-likelihood augmented by an $\ell_1$ penalty. The introduction of the $\ell_1$ penalty favors a parsimonious estimator of the regression coefficients, leading to convenient variable selection. The distribution of the least-squares estimator of a coefficient within the selected model, after conditioning on the selection event as well as the signs of the selected variables and nuisance parameters, is a truncated univariate Gaussian distribution \citep{lee2016exact}. Therefore, exact post-selection inference is possible in this case. 

However, this approach tends to yield excessively long confidence intervals due to the over-conditioning, which leaves little information for inference. In fact, it has been shown that this type of confidence interval has infinite expected length \citep{kivaranovic2021length}.  To address this issue, randomized versions of the lasso have been proposed to boost the inferential power. For example, \cite{tian2018selective} propose to add noise to the response vector and run the lasso on the noisy response. Another approach, known as data carving \citep{fithian2014optimal}, uses a subset of data for model selection. The two approaches can be shown to be asymptotically equivalent under certain conditions and we refer to both as the randomized lasso. In this situation, the randomness in the selection stage effectively smooths the boundary of the selection event. Consequently, the resulting conditional distribution is no longer a hard-truncated normal distribution, but can be viewed as a soft-truncated normal due to the marginalization over the randomness involved in selection. To conduct inference based on this distribution, prior works often resort to Markov chain Monte Carlo (MCMC) sampling, which is computationally intensive and might suffer from slow mixing, potentially resulting in unreliable inference.

A notable observation is that the lasso, along with several other model selection algorithms, gives rise to selection events that can be delineated as polyhedra when properly conditioned. Such methods, encompassing the lasso \citep{lee2016exact}, elastic net \citep{zou2005regularization}, square-root lasso \citep{belloni2011square,tian2018selective}, forward stepwise regression, least angle regression \citep{tibshirani2016exact}, SLOPE \citep{bogdan2015slope}, and the Benjamini-Hochberg procedure \citep{benjamini1995controlling,reid2017post}, fall under the category of polyhedral model selection. When the selection event is characterized by a polyhedron, it opens the door to utilizing more specialized sampling techniques tailored to this structure, surpassing the capabilities of generic MCMC algorithms. 

Thus, our first contribution is to introduce a more efficient sampling method designed specifically for conducting inference after polyhedral model selection. The method relies on the classic separation-of-variable method (SOV \citep{genz1992numerical}), which transforms the variables constrained by a polyhedron into variables within a unit cube. Subsequently, sampling from the unit cube can be executed with enhanced efficiency by leveraging conventional numerical integration methods, such as randomized quasi-Monte Carlo (QMC). The obtained p-values are highly accurate and equipped with error estimates. Furthermore, we develop a method that uses a single batch of QMC samples to construct confidence intervals for all selected variables, substantially reducing the computational workload.

Some recently proposed inference methods for the randomized lasso bypass the need for sampling. For instance, \citet{panigrahi2022approximate} present the method of approximate selective maximum likelihood estimator (MLE), which is based on the approximate normality of the MLE of the selection-adjusted likelihood. Their method computes the approximate MLE and uses the corresponding Fisher information matrix to construct Wald-type confidence intervals. Although it offers computational efficiency, this approach approximates both the MLE and its corresponding Fisher information matrix, relying on a Laplace approximation of the selection event probability. Consequently, this method is not reliable when the Laplace approximation falters.

To address this issue, our second contribution is an optimization algorithm that directly maximizes the selective likelihood without using any large-deviation type approximations. Our approach involves computing the gradient of the log-likelihood via the SOV technique and subsequently employing gradient ascent. Upon convergence, the SOV method is applied again to evaluate the Hessian at the maximum. Because the SOV method is highly accurate in estimating the gradients and Hessian, this provides a more dependable approach for conducting MLE-based inference after selection.

The remainder of the paper is structured as follows. In Section~\ref{sec: randomized lasso}, we introduce the background of the randomized lasso, and the two approaches for inference: one based on the cumulative distribution function (CDF)  of the conditional distribution and one based on selective MLE. We discuss some related work at the end. In Section~\ref{sec: sov}, we introduce the SOV method, which is employed to compute the p-value under the selection-adjusted distribution. Additionally, we propose several variance reduction techniques and present an algorithm designed to construct confidence intervals for all target parameters employing a single set of samples. In Section~\ref{sec: mle}, we elaborate on the utilization of the proposed method for optimizing the selective likelihood and conducting MLE-based inference. To demonstrate the effectiveness of the proposed method, we present numerical results in Section~\ref{sec: simulations}. Section~\ref{sec: conclusion} has our conclusions. Proofs and additional numerical results are in the Appendix. Code for the algorithm is accessible through the GitHub repository at \url{https://github.com/liusf15/selinf_sampler}.

\section{Inference after randomized lasso}
\label{sec: randomized lasso}

As mentioned earlier, incorporating randomness during the selection stage preserves more information for subsequent inference, thereby increasing the inferential power.  In this section, we delve into the randomized lasso problem and outline the framework for conducting conditional post-selection inference within this context.

\subsection{Randomized lasso and inference target}
Consider a dataset comprising $n$ data points $(\bfx_i,y_i)\in\R^p\times\R$, where $y_i$ is the response and $\bfx_i$ represents the potentially high-dimensional feature. These observations are organized into the response vector $Y$ of size $n$ and the design matrix $X$ of size $n\times p$. We assume a normal homoscedastic model $Y\sim\N(\bmu,\sigma^2 I)$ while leaving $\bmu\in\R^n$ unspecified. We assume $\sigma^2$ is known in this section.

In scenarios where not all $p$ features contribute meaningfully to predicting or interpreting the response, a model selection algorithm can be employed to identify a pertinent subset of features. For instance, the lasso \citep{tibshirani1996regression} solves a regularized optimization problem with the sparsity-inducing $\ell_1$ penalty. To incorporate randomness into the selection process, we solve the following randomized lasso problem
\begin{align}
\hat\bbeta^\lambda=\argmin_{\bbeta\in\R^p}\frac{1}{2}\|Y-X\bbeta\|_2^2+\lambda\|\bbeta\|_1 -\bomega\tran\bbeta,
\label{equ: randomized lasso}
\end{align}
where $\bomega\sim\N_p(\bzero, \Omega)$ is the randomization variable that is generated independently of the data, and $\lambda$ stands for the regularization parameter. This formulation can be shown to be asymptotically equivalent to the data carving, which uses a subset of the data for the lasso.
Let $M=\{j\in[p]:\; \hat\beta^\lambda_j\neq 0 \}$ denote the set of selected variables and let $d=|M|$ be the number of selected variables. We will assume the selected model has full rank such that $\text{rank}(X_M)=d$.

The inference target within the selected model is often nonstandard. For instance, in the submodel view \citep{berk2013valid}, the inference target is defined to be $X_M^\dagger \EE{Y}$, which is the projection of $\mu=\EE{Y}$ onto the column space spanned by the selected features $X_{M}$. Alternatively, the full model view chooses the inference target to be $\bbeta^f:=X^\dagger \EE{Y}$, which is well-defined only when $\text{rank}(X)=p$. Particularly, the parameters of interest are those $\beta^f_j$ for $j\in M$.


In either case, the inference target can be represented as $\bbeta_M=\calA_M\EE{Y}$ for some $d\times p$ matrix $\calA_M$ depending on $M$. Inference can then be based on
\begin{align}
\label{equ: A_M Y distribution}
\calA_M Y\sim\N(\bbeta_M, \sigma^2 \calA_M\calA_M\tran ).
\end{align}
In the submodel view, $\calA_M=X_M^\dagger$, resulting in $\calA_M Y\sim\N(\bbeta_M, \sigma^2 (X_M\tran X_M)^{-1} )$, where $\calA_M Y$ coincides with the least-squares estimator by regressing $Y$ onto the selected variables $X_M$. In the full model view, $\calA_M=J_MX^\dagger$, where $J_M\in\R^{d\times p}$ consists of the rows of the identity matrix $I_p$ with indices in $M$. 

Hereafter, we denote the test statistic as $\hat\bbeta_M=\calA_MY$ and the covariance matrix as $\Sigma=\sigma^2\calA_M\calA_M\tran$.
Thus, Equation~\eqref{equ: A_M Y distribution}, along with $\bomega\sim\N_p(\bzero,\Omega)$, leads to
\begin{align}
\label{equ: hat beta_M omega distribution}
\begin{pmatrix} \hat\bbeta_M \\ \bomega \end{pmatrix}
\sim\N_{d+p}\left(
\begin{pmatrix}\bbeta_M \\ \bzero \end{pmatrix}
,\begin{pmatrix}
\Sigma & \bzero \\ \bzero & \Omega
\end{pmatrix}\right).
\end{align}
Because $M$ is selected from the observed data, inference must be performed based on the above distribution while conditioning on the selection event.

\subsection{Selection-adjusted distribution}

To characterize the conditional distribution of $(\hat\bbeta_M,\bomega)\mid\{\whM=M\}$, we first study the selection event $\{\whM=M\}$. Here, we use $\whM$ to denote the random variable of the model selected by the lasso problem~\eqref{equ: randomized lasso}, and use $M$ to denote the realization of $\whM$ on the observed data. Recall that the Karush-Kuhn-Tucker (KKT) condition for problem~\eqref{equ: randomized lasso} is given by
\begin{align*}
&X\tran (X\hat\bbeta^\lambda-Y)+\bfs = \bomega\\
&\bfs_M=\lambda\cdot \sign(\hat\bbeta^\lambda_M),\;\; \|\bfs_{-M}\|_\infty\leq\lambda.
\end{align*}
Here, $\bfs$ denotes the subgradient of the penalty $\lambda\|\bbeta\|_1$ at the solution $\hat\bbeta^\lambda$.
By rearranging, we have
\begin{align*}
\bomega&= X\tran X_M \hat\bbeta_M^\lambda - X\tran X_M\hat\bbeta_M + X\tran(X_M\hat\bbeta_M-Y)+\bfs\\
&=X\tran X_MD\bfb  - X\tran X_M\hat\bbeta_M  + \bfr + \bfs,\numberthis\label{equ: kkt map}
\end{align*}
where $D=\diag(\sign(\hat\bbeta^\lambda_M))$ is the diagonal matrix of the signs of $\hat\bbeta^\lambda_M$, $\bfb=D\hat\bbeta^\lambda_M $ is the absolute value of $\hat\bbeta^\lambda_M$, and $\bfr=X\tran(X_M\hat\bbeta_M-Y)$ represents the residual. If we condition not only on the selection event but also on the sign of the lasso solution $\hat\bbeta^\lambda_M$ and the residual $\bfr$, then Equation~\eqref{equ: kkt map} provides a one-to-one linear mapping from $(\hat\bbeta_M,\bomega )$ to $(\hat\bbeta_M, \bfb, \bfs_{-M})$: 
\begin{align}
\bomega=Q_1 \hat\bbeta_M + Q_2\bfb +  \bfr+\bfs,
\label{equ: kkt map Q_1}
\end{align}
where $Q_1=-X\tran X_M$, $Q_2=X\tran X_MD$, $\bfr$, and $\bfs_M$ are all constants when conditioned on $\{\bfr,\; \bfs_{M} \}$. 
Recall that the unconditional distribution of $(\hat\bbeta_M,\bomega)$ is the product Gaussian distribution given in Equation~\eqref{equ: hat beta_M omega distribution}. Additionally, the KKT condition implies that $\bfb$ is componentwise positive and $\|\bfs_{-M}\|_\infty\leq\lambda$.
By applying this change-of-variable to the conditional distribution of $(\hat\bbeta_M,\bomega)\mid\{\bfr,\bfs_{-M} \}$, we obtain the conditional density of $(\hat\bbeta_M,\bfb,\bfs_{-M})\mid\{\bfr, \bfs_{M} \}$:
\begin{align*}
p(\hat\bbeta_M,\bfb,\bfs_{-M}\mid \{\bfr, \bfs_{M} \})\propto \varphi(\hat\bbeta_M;\bbeta_M, \Sigma )\cdot \varphi(Q_1 \hat\bbeta_M+Q_2 \bfb + \bfr+\bfs;\mathbf{0},\Omega  )\cdot \Indc{\bfb> 0,\|\bfs_{-M}\|_\infty\leq\lambda} .
\end{align*}
Lastly, we condition on $\bfs_{-M}$ and marginalize over $\bfb$ to obtain the conditional density of $\hat\bbeta_M\mid \{\bfr,\bfs \}$:
\begin{align}
p(\hat\bbeta_M\mid \{\bfr,\bfs \}) = \frac{\varphi(\hat\bbeta_M;\bbeta_M, \Sigma )\cdot \int_{\bfb\in\calO} \varphi(Q_1 \hat\bbeta_M+Q_2 \bfb + \bfr+\bfs;\mathbf{0},\Omega  )\rd \bfb }{\int_{\R^d}  \varphi(\hat\bbeta_M;\bbeta_M, \Sigma )\cdot \int_{\bfb\in\calO} \varphi(Q_1 \hat\bbeta_M+Q_2 \bfb + \bfr+\bfs;\mathbf{0},\Omega  )\rd \bfb \rd \hat\bbeta_M },
\label{equ: hat beta cond density}
\end{align}
where $\calO=\{\bfv\in\R^d: \bfv> 0 \}$ represents the positive orthant in $\R^d$. We will see later that, with certain choice of the covariance matrix $\Omega$ of the randomization variable, the above distribution is independent of $\bfr_{-M}$ and $\bfs_{-M}$. Thus conditioning on $\{\bfr_{-M}, \bfs_{-M}\}$ is not necessary.

In the following, we denote $\calE=\sigma(\{\bfr,\bfs \})$ as the $\sigma$-algebra generated by the random variables $\bfr$ and $\bfs$. Before describing methods for conducting inference for $\bbeta_M$ based on this conditional distribution, it is worth noting that data carving is asymptotically equivalent to the lasso with added noise as expressed in problem~\eqref{equ: randomized lasso}.
\begin{rmk}\label{rmk: carving}
Initially proposed in \cite{fithian2014optimal}, data carving refers to the approach that employs a subset of data for selection and uses the remaining information along with the hold-out data for inference. Specifically, this approach takes a subset of data $(X^{(1)},Y^{(1)})$ comprising $n_1<n$ observations and solves the lasso problem
\begin{align*}
\hat\bbeta^\lambda=\argmin_{\bbeta\in\R^p} \frac{1}{2 \rho}\| Y^{(1)} - X^{(1)}\bbeta\|_2^2 + \lambda\|\bbeta\|_1,
\end{align*}
where $\rho=n_1/n\in(0,1)$. This problem can be expressed in the form as Problem~\eqref{equ: randomized lasso}, with the randomization variable defined as
\begin{align*}
\bomega= - X\tran (Y-X\hat\bbeta^\lambda) + \frac1{\rho} X^{(1),\intercal}(Y^{(1)} - X^{(1)} \hat\bbeta^\lambda ),
\end{align*}
Moreover, as $n\goinf$ and $n_1/n\to\rho$ and with fixed $p$, $\bomega$ can be shown to have the asymptotic distribution $\N(\bzero,\frac{1-\rho}{\rho} \sigma^2 X\tran X )$. See \cite{markovic2016bootstrap} and \cite{liu2023selective} for the asymptotic justification in the context of linear and generalized linear models. Hence, data carving is asymptotically equivalent to the randomized lasso problem in \eqref{equ: randomized lasso} with $\Omega=\frac{1-\rho}{\rho}\sigma^2 X\tran X$.
\end{rmk}

\subsection{CDF-based inference}
\label{sec: cdf inference}

We focus on conducting inference for linear contrasts of the form $\bfeta\tran\bbeta_{M}$ for some $\bfeta\in\R^d$. For testing the null hypothesis $H_0:\bfeta\tran\bbeta_M=\theta$, a valid p-value can be obtained by considering the tail probability of the observed value of $\hat\theta:=\bfeta\tran\hat\bbeta_M$ within the conditional distribution given by Equation~\eqref{equ: hat beta cond density}. To eliminate the nuisance parameters in $\bbeta_M$, we condition further on $\bbeta^\perp:=\hat\bbeta_M - \bfc \hat\theta$, where $\bfc=\nu^{-1}\Sigma\bfeta $, $\nu=\bfeta\tran \Sigma \bfeta$. 

Let $\bft=\bfr+\bfs+Q_1\bbeta^\perp$, $\widetilde Q_1=Q_1\bfc$. Then the KKT map in Equation~\eqref{equ: kkt map Q_1} can be expressed as
\begin{align*}
\bomega=\wtQ_1 \hat\theta + Q_2\bfb + \bft.
\end{align*}
Similar to Equation~\eqref{equ: hat beta cond density}, the conditional density of $\hat\theta$ given $\{\calE,\;\bbeta^\perp \}$ can be expressed as
\begin{align*}
\frac{\varphi(\hat\theta;\theta,\nu)\cdot \int_{\bfb\in\calO} \varphi( \widetilde Q_1\hat\theta +Q_2\bfb +\bft;\mathbf{0},\Omega )\rd \bfb} 
{\int_{\R}\varphi(\hat\theta;\theta,\nu)\cdot \int_{\bfb\in\calO} \varphi( \widetilde Q_1\hat\theta +Q_2\bfb +\bft;\mathbf{0},\Omega )\rd \bfb\rd \hat\theta}.
\end{align*}
The CDF of the above distribution can serve as a one-sided p-value. The following proposition provides an expression of the CDF. Define $H=Q_2\tran\Omega^{-1}Q_2$, $\bfk=Q_2\tran\Omega^{-1}\bft$, $\tilde{\bfc}=D\bfc$.
\begin{proposition}[CDF of $\hat\theta\mid\{\calE,\bbeta^\perp\}$]\label{prop: cdf}
Under $H_0:\bfeta\tran\bbeta_M=\theta$, the CDF $F(x)$ of the distribution of $\hat\theta$  conditional on $\{\calE,\;\bbeta^\perp \}$ is given by
\begin{align}
\label{equ: cdf}
F(x)=\frac{\int_{\bfb\in\calO} \Phi(\frac{x - \mu_{\hat\theta}(\bfb) }{\sigma_{\hat\theta} } )\varphi(\bfb; \bmu_{\bfb},\Sigma_{\bfb} )\rd \bfb  }{\int_{\bfb\in\calO} \varphi(\bfb;\bmu_{\bfb},\Sigma_{\bfb} )\rd \bfb  },
\end{align}
where
\begin{equation*}
\begin{aligned}
&\Sigma_\bfb(\bfeta)=H^{-1}+\nu \tilde{\bfc} \tilde{\bfc}\tran,\\
&\sigma_{\hat\theta}^2=(\nu^{-1} + \tilde{\bfc}\tran H\tilde{\bfc} )^{-1}, 
\end{aligned}\qquad
\begin{aligned}
&\Sigma_\bfb(\bfeta)^{-1}\bmu_\bfb(\bfeta)=-\bfk + \sigma_{\hat\theta}^2 H\tilde{\bfc} (\nu^{-1}\theta + \tilde{\bfc}\tran \bfk ) ,\\
&\sigma_{\hat\theta}^{-2}\mu_{\hat\theta}(\bfb)=\nu^{-1}\theta + \tilde{\bfc}\tran \bfk + \tilde{\bfc}\tran H\bfb .
\end{aligned}
\end{equation*}
\end{proposition}
See Appendix~\ref{proof: prop cdf} for the derivation. The derivation involves decomposing the joint distribution of $(\hat\theta, \bfb)$ into the conditional distribution of $\hat\theta\mid\bfb$, which is $\N(\mu_{\hat\theta}(\bfb), \sigma_{\hat\theta}^2 )$, and the marginal distribution of $\bfb$, which is $\N(\bmu_\bfb(\bfeta),\Sigma_\bfb(\bfeta))$.

Note that the conditional variance $\sigma_{\hat\theta}^2$  and conditional mean $\mu_{\hat\theta}(\bfb)$ of $\hat\theta$ given $\bfb$ also depend on $\bfeta$, but we omit this dependence for simplicity of notation. Moreover, the expressions of $\bmu_\bfb(\bfeta)$ and $\mu_{\hat\theta}(\bfb)$ depend on $\bft$, which encompasses the random variables that are conditioned on. These expressions also depend on the value of $\theta$. However, the variances $\Sigma_\bfb(\bfeta)$ and $\sigma_{\hat\theta}^2$ remain unaffected by $\bft$ or $\theta$.

Recall from Remark~\ref{rmk: carving} that data carving can be viewed as Problem~\eqref{equ: randomized lasso} with the covariance of $\bomega$ equal to $\frac{1-\rho}{\rho}\sigma^2 X\tran X$. In this special case, the CDF $F(x)$ in Proposition~\ref{prop: cdf} does not depend on $\bft_{-M}$ and the expression can be simplified.
\begin{corollary}[$\Omega=\kappa^{-1}\cdot \sigma^2 X\tran X$]
\label{coro: carving Sigma_omega}
If $\Omega=\kappa^{-1}\cdot \sigma^2 X\tran X $, then {\sloppy $\bfk=Q_2\tran\Omega^{-1}\bft=(\kappa/\sigma^2)D\bft_M$}. Consequently, the CDF in Proposition~\ref{prop: cdf} does not depend on $\bft_{-M}$. Moreover, we have the simplified expressions
\begin{equation*}
\begin{aligned}
&\Sigma_\bfb(\bfeta)=H^{-1}+\nu\tilde{\bfc}\tilde{\bfc}\tran, \\
&\sigma_{\hat\theta}^2=\frac{\nu}{1+\kappa},
\end{aligned}\qquad
\begin{aligned}
&\bmu_\bfb(\bfeta)=-H^{-1}\bfk + \tilde{\bfc}\theta,  \\
&\mu_{\hat\theta}(\bfb)=\frac{\nu}{1+\kappa}\Big(\frac{\theta}{\nu}+\frac{\kappa}{\sigma^2} c\tran \bft_M + \frac{\kappa}{\nu}\eta\tran D\bfb \Big).
\end{aligned}
\end{equation*}
\end{corollary}

This corollary indicates that when $\Omega$ is a multiple of $X\tran X$, as is the case for data carving, the conditional distribution of $\hat{\theta}$ becomes independent of $\bfs_{-M}$ and $\bfr_{-M}$. In essence, it is equivalent to the distribution that is not conditioned on $\bfs_{-M}$ and $\bfr_{-M}$. Consequently, choosing the randomization covariance matrix $\Omega$ to be a multiple of $X\tran X$ effectively reduces the conditioning set from  $\{\bfs, \bfr, \hat\bbeta^\perp \}$ to $\{\bfs_M, \bfr_M, \hat\bbeta^\perp \}$.

The CDF in Equation~\eqref{equ: cdf}, evaluated at the observed value of $\hat\theta$, is a valid p-value since $F(\hat\theta)\sim \unif([0,1])$ conditional on $\{\whM=M\}$. Similarly, $1-F(\hat\theta)$ and $2\cdot \min\{F(\hat\theta), 1-F(\hat\theta) \}$ are also uniformly distributed when conditioned on $\{\whM=M\}$. Consequently, they can be used to test the hypothesis $H_0:\bfeta\tran\bbeta_M=\theta$. However, it is challenging to evaluate the CDF in Equation~\eqref{equ: cdf} due to the integral over $\bfb\in\calO$. In fact, the CDF can be equivalently expressed as $F(x)=\EE{\Phi(\frac{x - \mu_{\hat\theta}(\bfb) }{\sigma_{\hat\theta}} )}$, where the expectation is taken over $\bfb\sim \N(\bmu_\bfb,\Sigma_\bfb)\mid _\calO$, the normal distribution truncated to the positive orthant. The main objective of this study is to provide a numerical integration algorithm to compute this expectation efficiently.

\subsection{MLE-based inference}
\label{sec: intro mle}
Aside from the CDF-based approach, another method for inference hinges on the approximate normality of the MLE of the selective likelihood. The following proposition provides an expression of the selective likelihood, derived from the selection-adjusted distribution of $\hat\bbeta_M$.
\begin{proposition}[Selective likelihood]\label{prop: selective likelihood}
The conditional density in Equation~\eqref{equ: hat beta cond density} yields the selective likelihood up to a constant:
\begin{align*}
\ell(\bbeta_M)=\frac{\varphi(\hat\bbeta_M; \bbeta_M,\Sigma )}{\int_{\bfb\in\calO} \varphi(\bfb;\bmu_{\bfb},\Sigma_{\bfb})\rd \bfb  },
\end{align*}
where 
\begin{align*}
\Sigma_\bfb=H^{-1} + D\Sigma D,\quad \bmu_\bfb(\bbeta_M)=D \bbeta_M - H^{-1}Q_2\tran\Omega^{-1}(\bfr+\bfs).
\end{align*}
\end{proposition}
The numerator of the selective likelihood corresponds to the unconditional density of $\hat\bbeta_M$, which mirrors the standard likelihood without any adjustments. The denominator arises from the normalizing constant of the density in Equation~\eqref{equ: hat beta cond density}, which is equal to the probability of selecting the model. See the derivation in Appendix~\ref{prf: selective likelihood}.

The selective MLE $\hat\bbeta^{\text{sMLE}}$ is thus the maximum of the likelihood function $\ell(\bbeta_M)$.
\citet{panigrahi2022approximate} demonstrate that the distribution of $\betamle$ can be approximated by the normal distribution $\N(\bbeta_M, (I^{\text{sMLE}})^{-1} )$, where $I^{\text{sMLE}}=-\nabla^2 \log \ell(\betamle)$ is the Hessian of the negative logarithm of the selective likelihood at $\hat\bbeta^{\text{sMLE}}$. Given $\betamle$ and $I^{\text{sMLE}}$, a level-($1-\alpha$) confidence interval for $\bfeta\tran\bbeta_{M}$ can be constructed as
\[
\bfeta\tran\hat\bbeta_{M}\pm q_{1-\alpha/2} \sqrt{\bfeta\tran (I^{\text{sMLE}})^{-1} \bfeta },
\]
where $q_{1-\alpha/2}=\Phi^{-1}(1-\alpha/2)$ is the $(1-\alpha/2)$ quantile of the standard normal distribution. The advantage of this approach is that it avoids the need to condition on $\bbeta^\perp$ to eliminate nuisance parameters, which is required in the CDF-based inference. A limitation of this method is that the distribution of $\hat\bbeta^{\text{sMLE}}$ is approximated as a Gaussian distribution, rather than being exactly Gaussian. For example, the Gaussian approximation might perform poorly in cases where the randomization level is weak, as shown by \cite{panigrahi2022approximate}.

\cite{panigrahi2022approximate} propose an approximate method to find the selective MLE and its corresponding Fisher information. Their method relies on a large deviation approximation of the selection probability, which can lead to unreliable results if this approximation is not accurate. In this work, we provide an algorithm to compute the MLE and Fisher information more precisely.

\subsection{Related work}

In the literature of conditional post-selection inference, the conditional distributions similar to the one in Equation~\eqref{equ: hat beta cond density} are often handled by MCMC sampling, such as Gibbs sampling \citep{tian2016magic}, hit-and-run \citep{belisle1993hit,fithian2014optimal,tian2018selective}, and projected Langevin dynamics \citep{markovic2016bootstrap}. The substantial computational demands of these MCMC-based methods is the main motivation for the development of the more efficient sampling algorithm in this work.

Several sampling-free methods have been specifically designed for the randomized lasso. The approximate MLE method by \cite{panigrahi2022approximate} mentioned earlier is one example. \citet{panigrahi2022exact} propose a method that involves further conditioning to make the conditional distribution tractable, enabling the derivation of exact p-values and confidence intervals. However, this ``over-conditioning" could potentially diminish statistical power. A comparative evaluation of these methods is conducted in simulation in Section~\ref{sec: simulations}. 

A common criticism of randomized selection algorithms resolves around the variability in the selected model due to different realizations of randomness. To address this concern, \cite{schultheiss2021multicarving} propose to perform multicarving across multiple partitions of the data to enhance robustness and reproducibility. However, this approach requires MCMC sampling for every single carving and every single parameter, thereby restricting its practicality. In this context, our proposed method can play a crucial role in reducing the computational demands of multicarving.

Post-selection inference with more complex data or selection algorithms is an active area of research. \cite{liu2023selective} develop an algorithm for post-selection inference with distributed data, where model selection is performed locally and only summary statistics are communicated to deliver inference for the aggregated model. \cite{liu2022black} propose a generic approach for inference after a general model selection. Their approach heavily relies on the assumption that the selection procedure can be repeatedly executed on bootstrapped datasets in order to acquire knowledge about the selection event.

Beyond the conditional approach, the PoSI framework introduced by \cite{berk2013valid} and extended by \cite{bachoc2019valid,bachoc2020uniformly,kuchibhotla2020valid} provides simultaneous inference that guarantees validity across all selection procedures. Consequently, these methods can be conservative due to their worst-case guarantee.

More recently, \cite{rasines2021splitting} propose a splitting strategy that mimics sample splitting, but the information splitting is conducted via added Gaussian noise. This method, termed the $(U,V)$ decomposition, involves generating a Gaussian vector $W$ such that $U=Y+\gamma W$ is independent of $V=Y-\gamma^{-1}W$. Model selection is carried out on $U$ while inference is conducted on $V$. Due to the independence between $U$ and $V$, the inference is free of selection bias. \cite{leiner2021data} generalize this approach to settings where the data is not necessarily Gaussian and name their approach ``data fission". 


\section{Separation-of-variable method}
\label{sec: sov}

In this section, we will describe the proposed method for computing the CDF $F(x)$ for some fixed $x\in\R$ in Equation~\eqref{equ: cdf}. For simplicity, we will denote $\bmu=\bmu_{\bfb}$, $\Sigma=\Sigma_{\bfb}$, as well as
\[
\bfg_1=-\sigma_{\hat\theta} H\tilde{\bfc},\quad  g_2=\frac{x}{\sigma_{\hat\theta}}- \sigma_{\hat\theta}(\nu^{-1}\theta + \tilde{\bfc}\tran\bfk )
\]
such that $\bfg_1\tran\bfb + g_2=\frac{1}{\sigma_{\hat\theta}} ( x - \mu_{\hat\theta}(\bfb) ) $. With this notation, $F(x)$ can be expressed as
\begin{align}
F(x)=\frac{\int_{\bfb\in\calO} \Phi(\bfg_1\tran \bfb + g_2 ) \varphi(\bfb;\bmu,\Sigma) \rd \bfb } {\int_{\bfb\in\calO}\varphi(\bfb;\bmu, \Sigma) \rd \bfb}.
\label{equ: pivot 2}
\end{align}

We begin by observing that the denominator of Equation~\eqref{equ: pivot 2} is equal to the orthant probability of the distribution $\N(\bmu,\Sigma)$. The most widely-used method to compute this orthant probability is the separation-of-variable (SOV) technique introduced by \cite{genz1992numerical}. The SOV method starts by reparameterizing $\bfb$ as $\bfb=L\bfz+\bmu$, where $L$ is the Cholesky decomposition of $\Sigma$ and $\bfz\sim\N(0,I)\mid_{L\bfz+\bmu\in\calO}$. This enables us to express $F(x)$ equivalently as
\begin{align*}
F(x)=\frac{\int_{L\bfz+\bmu\in\calO} \Phi(\tilde{\bfg}_1\tran \bfz + \tg_2 ) \varphi(\bfz) \rd \bfz } {\int_{L\bfz+\bmu\in\calO}\varphi(\bfz) \rd \bfz},\;\text{ where }\tilde{\bfg}_1=L\tran \bfg_1,\; \tg_2=g_2+\bfg_1\tran\bmu.
\end{align*}

Because $L$ is lower-triangular and has positive diagonals, i.e. $L_{kk}>0$ for all $k$, the constraint of $L\bfz+\bmu\in\calO$, which is $\mu_k+\sum_{j=1}^{k}L_{kj}z_j> 0$ $\forall\, k\in[d]$, can be sequentially expressed out for each variable $k=1,2,\ldots,d$ as follows:
\begin{align*}
z_k\geq \frac{-\mu_k-\sum_{j=1}^{k-1}L_{kj}z_j }{L_{kk}}=:a_k(\bfz_{1:k-1}).
\end{align*}
The value of $a_k$ depends on the realization of the previous $k-1$ variables $\bfz_{1:k-1}$, and serves as the lower bound of $z_k$ in the spherical Gaussian space.
Consequently, the denominator is equal to
\begin{align*}
\int_{L\bfz+\bmu\in\calO}\varphi(\bfz) \rd \bfz&=\int_{z_1\geq a_1} \varphi(z_1)\int_{z_2\geq a_2(z_1)} \varphi(z_2)\cdots \int_{z_d\geq a_d(\bfz_{1:d-1})} \varphi(z_d)\rd \bfz  \\
&=\int_{\Phi(a_1)}^1 \int_{\Phi(a_2)}^{1}\cdots  \int_{\Phi(a_d)}^1 \rd \bfv \qquad (\text{where }v_k=\Phi(z_k)) \\
&{=}\int_{[0,1]^d} \prod_{k=1}^d (1-\Phi(a_k)) \rd \bfu \qquad  (\text{where } v_k=\Phi(a_k) +  (1-\Phi(a_k)) u_k )
\end{align*}
where the second equality applies the change-of-variable $z_k=\Phi^{-1}(v_k)$ and the third equality shifts and scales the variables $v_k$ so that the integral is over the unit cube $[0,1]^d$. These transformations establish the relationships among $\bfu,\bfa,\bfz$ as
\begin{align}
a_k=\frac{-\mu_k -\sum_{j=1}^{k-1}L_{kj}z_j }{L_{kk}},\quad z_k=\Phi^{-1}(\Phi(a_k) + u_k(1-\Phi(a_k)) ),\quad \forall\, 1\leq k\leq d.
\label{equ: sov a_k, z_k}
\end{align}
Throughout the rest of the paper, these dependencies should always be understood, even if not explicitly stated. In addition, we always assume $\bfb=L\bfz+\bmu$.

Analogously, the numerator $\int_{L\bfz+\bmu\in\calO} \Phi(\tilde{\bfg}_1\tran \bfz + \tg_2 ) \varphi(\bfz) \rd \bfz$ can also be expressed as the integral
$$
\int_{[0,1]^d}\Phi(\tilde{\bfg}_1\tran \bfz + \tg_2) \prod_{k=1}^d (1-\Phi(a_k)) \rd \bfu,
$$
where $\bfz,\bfa$ also satisfy Equation~\eqref{equ: sov a_k, z_k}. We define the SOV weight $w(\bfu)$ as the function
\begin{align}
\label{equ: define sov weight}
w(\bfu)=\prod_{k=1}^d(1-\Phi(a_k)),
\end{align}
with the understanding that the dependence of $a_k$ on $\bfu$ is implied by Equation~\eqref{equ: sov a_k, z_k}. This leads us to the expression
\begin{align}
F(x)=\frac{ \int_{[0,1]^d}\Phi(\tilde{\bfg}_1\tran \bfz + \tg_2) w(\bfu) \rd \bfu}{\int_{[0,1]^d} w(\bfu) \rd \bfu}.
\label{equ: pivot transformed}
\end{align}

This expression allows us to estimate $F(x)$ by generating uniform samples $\bfu^{(1)},\ldots,\bfu^{(N)}$ in the unit cube $[0,1]^d$ and computing
\begin{align}
\widehat{F(x)}=\frac{\frac1N\sum_{i=1}^N \Phi(\tilde{\bfg}_1\tran \bfz^{(i)}+ \tg_2 ) w(\bfu^{(i)}) }{\frac1N\sum_{i=1}^N w(\bfu^{(i)}) },
\label{equ: sov estimator}
\end{align}
where each $\bfz^{(i)}$ is determined by $\bfu^{(i)}$ via Equation~\eqref{equ: sov a_k, z_k}. It is worth noting that this estimator is equivalent to the self-normalized importance sampling (SNIS) estimator using proposal density
\[
\prod_{k=1}^d \frac{\varphi(z_k)\Indc{z_k\geq a_k(\bfz_{1:k-1}) }}{1-\Phi(a_k(\bfz_{1:k-1}) ) },
\]
since the ratio between $\varphi(\bfz)\Indc{L\bfz+\bmu\in\calO}$ and the above proposal density is exactly equal to the SOV weight $w(\bfu)$.

\subsection{Variance reduction}

In order to enhance the precision of the Monte Carlo estimator in Equation~\eqref{equ: sov estimator}, we can employ several variance reduction techniques.

\paragraph*{Conditional Monte Carlo}

First, we observe that the denominator of the quantity given in Equation~\eqref{equ: pivot transformed} is in fact a $(d-1)$-dimensional integral. This is because $a_k$ only depends on $\bfu_{1:k-1}$, thus the SOV weight $w(\bfu)=\prod_{k=1}^d (1-\Phi(a_k))$ does not depend on $u_d$. If the numerator can also be evaluated using $\bfu_{1:d-1}$ only, then it suffices to sample from the $(d-1)$-dimensional unit cube instead of the $d$-dimensional unit cube.

To achieve this, we will integrate out $u_d$ from the integrand in the numerator exactly. This technique is known as conditional Monte Carlo, or pre-integration, or Rao-Blackwellization \citep{blackwell1947conditional,casella1996rao}. By integrating out some variable in a closed form, we always reduce the Monte Carlo variance of the estimator.
Note that the integrand in the numerator is given by $\Phi(\tilde{\bfg}_1\tran \bfz + \tg_2)\prod_{k=1}^d (1-\Phi(a_k))$. Integrating out $u_d$ can be carried out as follows:
\begin{align*}
&\int_0^1 \Phi(\tg_{1,d} z_d +  \tilde{\bfg}_{1,-d}\tran \bfz_{-d} + \tg_2)\prod_{k=1}^d (1-\Phi(a_k)) \rd u_d\\
&\quad\qquad=\prod_{k=1}^{d-1} (1-\Phi(a_k)) \cdot \int_{a_d }^{\infty} \Phi(\tg_{1,d} z_d +  \tilde{\bfg}_{1,-d}\tran \bfz_{-d} + \tg_2)\varphi(z_d)\rd z_d.
\end{align*}
The above univariate integral can be expressed as the CDF of a bivariate normal distribution, which can be evaluated efficiently with high precision based on Owen's T function \citep{patefield2000fast}.

\paragraph*{Variable reordering} 

The SOV estimator's performance depends on the arrangement of the variables. As noted in \cite{genz1992numerical}, rearranging variables might lead to significant error reduction when computing the Gaussian orthant probabilities. \citet{gibson1994monte} introduced a heuristic method to reorder the variables so that the innermost integrals have the largest expected values. We apply the Gibson ordering to reorder the variables before employing the SOV method to compute the integrals.


\paragraph*{Quasi-Monte Carlo}

The integrands $w(\bfu)$ and $\Phi(\tilde{\bfg}_1\tran \bfz+\tg_2 ) w(\bfu)$ are bounded and smooth on the unit cube, making them highly suitable for quasi-Monte Carlo (QMC) sampling. If the points $\bfu^{(i)}$ ($1\leq i\leq N$) are sampled uniformly and independently in the unit cube, then the estimator in Equation~\eqref{equ: sov estimator} has a probabilistic error rate of $O_p(N^{-1/2})$ for both the numerator and the denominator separately. This error rate can be substantially improved by adopting QMC or randomized QMC (RQMC) methods. QMC points are chosen strategically and deterministically to cover the unit cube more evenly than i.i.d. Monte Carlo does. For integrands of bounded variation in the sense of Hardy-Krause, QMC achieves an error rate of $O(N^{-1+\delta})$ for any $\delta>0$, where $N^\delta$ hides the the log term $(\log N)^d$ \citep{niederreiter1992random}.


However, this error rate is obtained by a worst-case upper bound and is often too conservative to be useful as an error estimate. Moreover, it does not apply for functions with unbounded Hardy-Krause variation. One remedy to these issues is to apply randomization. In RQMC, the points $\bfu^{(i)}$ are uniformly distributed individually while collectively they have the low-discrepancy property of QMC. This ensures that the estimator of the integral is unbiased, and its standard error can be estimated by independent replicates. Randomization can be achieved by adding a random shift to lattice rules or by randomly scrambling the digits of digital nets. See \cite{l2002recent} for a review of RQMC. 

In this work, we propose to use scrambled Sobol' points \citep{sobo:1967:russ,rtms}, a particular type of RQMC points. These points are construction-free, thereby eliminating the need for case-by-case constructions as required by lattice rules. For sufficiently smooth integrands, it can achieve an error rate of $O(N^{-3/2+\delta})$ \citep{snetvar,owen1997scrambled}. In addition, it has been observed by \cite{hong2003algorithm} that scrambled Sobol' points perform empirically better than competing methods for this particular problem of computing multivariate normal probabilities.

\subsection{Confidence intervals}
\label{sec: CI sov}
In practice, there might be a need to test the hypothesis $\bfeta\tran\bbeta_M=\theta$ for a range of different values of $\theta$ and $\bfeta$. For instance, we might want to test $\beta_{M,j}=0$ for all $j\in M$. Moreover, constructing a confidence interval for $\bfeta\tran\bbeta_M$ involves inverting the hypothesis testing, requiring the computation of the pivotal quantity $F(x)$ defined in Proposition~\ref{prop: cdf} for various values of $\theta$. While we can apply the SOV method to compute p-values for each individual hypothesis separately, we offer an approach that facilitates testing the hypothesis $\bfeta\tran\bbeta_M=\theta$ for any $\bfeta$ and $\theta$ using just a single batch of RQMC samples.

Recall that when testing $\bfeta\tran\bbeta_M=\theta$, we require sampling $\bfb$ from $\N(\bmu_{\bfb}(\bfeta,\theta), \Sigma_{\bfb}(\bfeta) )\mid_{\calO} $, where
\begin{align*}
\Sigma_{\bfb}(\bfeta)^{-1}&=H-\sigma_{\hat\theta}^2 H\tilde{\bfc} \tilde{\bfc}\tran H,\\
\Sigma_{\bfb}(\bfeta)^{-1} \bmu_{\bfb}(\bfeta,\theta)&=-\bfk + \sigma_{\hat\theta}^2 H\tilde{\bfc}(\nu^{-1}\theta + \tilde{\bfc}\tran \bfk).
\end{align*}
according to Proposition~\ref{prop: cdf}. Here the notation emphasizes the dependence of $\bmu_{\bfb}$ and $\Sigma_\bfb$ on $\bfeta$ and $\theta$.

Rather than sampling from $\N(\bmu_{\bfb}(\bfeta, \theta), \Sigma_\bfb(\bfeta))\mid_{\calO}$, which depends on $(\bfeta,\theta)$, we propose to sample from the distribution $ \N(\bar\bmu, \bar\Sigma )\indc{\calO} $, where
\begin{align*}
\bar\Sigma^{-1}&=H,\\
\bar\Sigma^{-1}\bar\bmu&=-Q_2\tran\Omega^{-1}(\bfr+\bfs - X\tran X_M\hat\bbeta_M ).
\numberthis\label{equ: bar mu Sigma}
\end{align*}
This distribution is also supported on the orthant $\calO$ and is independent of $(\bfeta,\theta)$. The underlying rationale for this choice is to achieve a proximity between  $(\bar\Sigma^{-1},\bar\Sigma^{-1}\bar\bmu)$ and $(\Sigma_{\bfb}(\bfeta)^{-1}, \Sigma_{\bfb}(\bfeta)^{-1}\bmu_{\bfb}(\bfeta,\theta))$ across different values of $\bfeta$ and $\theta$.

To evaluate integrals w.r.t. the target distribution $\N(\bmu_{\bfb}(\bfeta, \theta), \Sigma_\bfb(\bfeta))\mid_{\calO}$, we apply importance weighting. The expression of the importance weight is given in the following lemma.
\begin{lemma}[Importance weight relative to $\N(\bar{\bmu}, \bar\Sigma)\mid_{\calO}$]\label{lem: importance weight}
Given $(\bfeta,\theta)$, let $\bfk$, $\tilde{\bfc}$ be as defined in Section~\ref{sec: cdf inference}.
Define $\Delta=\sigma_{\hat\theta} (\nu^{-1}\theta + \tilde{\bfc}\tran\bfk-\frac{\hat\theta}{\sigma_{\hat\theta}^2} )$ and $\btau=\sigma_{\hat\theta} H\tilde{\bfc}$. 
Then the importance weight $\frac{\varphi(\bfb;\bmu_\bfb(\bfeta,\theta),\Sigma_\bfb(\bfeta) )}{\varphi(\bfb;\bar\bmu,\bar\Sigma ) }\indc{\bfb\in\calO} $ is proportional to
\begin{align}\label{equ: define bar w}
\bar w(\bfb)&:=\Exp{\frac{1}{2}(\bfb\tran \btau)^2 +  \Delta (\bfb\tran \btau)  } \indc{\bfb\in\calO}.
\end{align}
\end{lemma}
See the derivation in Appendix~\ref{prf: lem importance weight}.
Notably, $\bar w(\bfb)$ is a function of the inner product $\bfb\tran\btau$. Thus, the computation of the importance weight requires only the evaluation of this vector-vector product, eliminating the need for matrix-vector products. Furthermore, this implies that the importance sampling effectively operates as a one-dimensional importance sampling, thereby circumventing the potential challenge related to products of weights in high dimensions.



Now we summarize the entire procedure. We first generate $N$ samples $\{\bfu^{(i)}\}_{1\leq i\leq N}$ within the unit cube $[0,1]^d$ by RQMC. Then we apply the SOV method to obtain samples $\{\bfb^{(i)}\}_{1\leq i\leq N}$ from $\N(\bar\bmu,\bar\Sigma)\mid_{\calO}$ with the associated SOV weights $\{w(\bfu^{(i)})\}_{1\leq i\leq N}$. For each sample $\bfb^{(i)}$, we calculate the importance weight $\bar w(\bfb^{(i)})$ given by Equation~\eqref{equ: define bar w} and then set $w^{(i)}=w(\bfu^{(i)})\cdot \bar w(\bfb^{(i)}) $. Then the weighted samples $\bfb^{(i)}$, weighted by $w^{(i)}$, are effectively drawn from $\N(\bmu_{\bfb}(\bfeta,\theta), \Sigma_{\bfb}(\bfeta) )\mid_{\calO}$.  For a particular pair of $\bfeta$ and $\theta$, an estimate of the CDF $F(x)$ defined in Equation~\eqref{equ: cdf} can be constructed similarly as Equation \eqref{equ: sov estimator}, with the weights being $w^{(i)}$. A valid one-sided p-value for testing $\bfeta\tran\bbeta_M=\theta$ is obtained by evaluating $F(x)$ (or $1-F(x)$) at the observed value of $\bfeta\tran\hat\bbeta_M$.

To construct a confidence for $\bfeta\tran\bbeta_M$, we need to invert the hypothesis tests for $\bfeta\tran\bbeta_M=\theta$.  To achieve this, we vary $\theta$ across a grid and form the interval of those values of $\theta$ for which the corresponding hypothesis is not rejected. For instance, we can take a grid within the interval $[\hat\beta_j^{(2)} -10s_j, \hat\beta_j^{(2)} + 10s_j]$, where $\hat{\bbeta}^{(2)}$ is the unbiased estimator of $\bbeta_M$ computed using the hold-out data only, and $s_j$ is the corresponding standard error. Hence, the interval $[\hat\beta_j^{(2)} -10s_j, \hat\beta_j^{(2)} + 10s_j]$ most likely covers all the $\theta$ that will not be rejected.
The whole procedure is summarized in Algorithm~\ref{algo}.
\begin{algorithm}
\caption{Confidence intervals for the components of $\bbeta_M$ using the SOV method}
\label{algo}
\begin{algorithmic}

\State Compute $\bar\bmu$ and $\bar\Sigma$ defined in Equation~\eqref{equ: bar mu Sigma}.
\State Apply the SOV method with RQMC to get $N$ weighted samples $\{(\bfb^{(i)}, w(\bfu^{(i)}))\}_{i=1}^N $ from $\N(\bar\mu,\bar\Sigma)\mid_{\calO}$.

\For{$j\gets 1,\ldots,d$}
\State Let $\bfeta=\bfe_j$ and compute $\bfk, \tilde{\bfc}$ as defined in Section~\ref{sec: cdf inference}. Compute $\sigma_{\hat\theta}^2 $ as in Proposition~\ref{prop: cdf}.

\State Let $\hat{\bbeta}^{(2)}$ be the least-squares estimator of $\bbeta_M$ using the hold-out data and let 
\[s_j=\sigma\sqrt{[(X^{(2),\intercal}_M X^{(2)}_M)^{-1}]_{jj} }.\]
\State Let $\calG$ be a grid on $[\hat\beta_j^{(2)} -10s_j, \hat\beta_j^{(2)} + 10s_j ]$ 
\For{$\theta$ in $\calG$ }
    \State Compute $\mu_{\hat\theta}(\bfb^{(i)})$ for $1\leq i\leq N$ as in Proposition~\ref{prop: cdf} with the specific value of $\theta$.
    \State Compute $w^{(i)}=w(\bfu^{(i)})\cdot \bar w(\bfb^{(i)})$, where $\bar w$ is defined in \eqref{equ: define bar w}.

    \State Compute the p-value by
    \begin{align*}
    p(\theta)=\frac{\sum_{i=1}^N \Phi(\frac{\hat\bbeta_{M,j} - \mu_{\hat\theta}(\bfb^{(i)}) }{\sigma_{\hat\theta}} ) w^{(i)} }{\sum_{i=1}^N w^{(i)} },
    \end{align*} 
    \State or $1-p(\theta)$ or $2\cdot \min\{p(\theta) ,1-p(\theta)\}$

\EndFor
    \State Form the confidence interval for $\beta_{M,j}$ as $\{\theta\in\calG: p(\theta)\geq1-\alpha \}$.
\EndFor
\Ensure{Level-$(1-\alpha)$ confidence intervals for $\beta_{M,j}$ ($1\leq j\leq d$) }

\end{algorithmic}
\end{algorithm}

\section{MLE-based inference}
\label{sec: mle}

This section describes how to maximize the selective likelihood given in Proposition~\ref{prop: selective likelihood} and conduct inference based on the MLE. The negative logarithm of the selective likelihood is equal to
\begin{align*}
 -\log\ell(\bbeta_M)=\frac12(\hat\bbeta_M-\bbeta_M )\tran \Sigma^{-1} (\hat\bbeta_M-\bbeta_M) + \log \int_{\bfb\in\calO} \varphi(\bfb;\bmu_{\bfb}(\bbeta_M),\Sigma_{\bfb})\rd \bfb,
\end{align*}
where $\bmu_{\bfb}(\bbeta_M)$ and $\Sigma_{\bfb}$ are given in Proposition~\ref{prop: selective likelihood}. 

We propose to run gradient descent to minimize the negative log-likelihood $-\log\ell(\bbeta_M)$. A challenge arises due to the presence of the orthant probability $\int_{\bfb\in\calO} \varphi(\bfb;\bmu_{\bfb}(\bbeta_M),\Sigma_{\bfb})\rd \bfb$ in the objective function. In our case, we need the gradient and Hessian of the log orthant probability w.r.t. $\bbeta_M$. Since the Jacobian of $\bmu_{\bfb}(\bbeta_M)$ w.r.t. $\bbeta_M$ is equal to $D$, it suffices if we can evaluate the gradient and Hessian of the log orthant probability w.r.t. $\bmu_\bfb$.
The next lemma provides expressions for the gradient and the Hessian of the log orthant probability.
\begin{lemma}[Gradient and Hessian of log orthant probability]\label{lemma: gradient orthant prob}
Let $h(\bmu):=\int_{\calO}\varphi(\bfb;\bmu,\Sigma)\rd \bfb$ denote the orthant probability as a function of $\bmu$.
Let $\tilde{\bmu}$ and $\widetilde{\Sigma}$ denote the mean and covariance matrix of the truncated normal distribution $\N(\bmu,\Sigma)\mid_{\calO}$. Then $\nabla_{\bmu} \log h(\bmu) $ and $\nabla^2_{\bmu} \log h(\bmu) $ have the expressions
\begin{align*}
\nabla_{\bmu} \log h(\bmu) &=\Sigma^{-1}(\tilde{\bmu} - \bmu ), \\
\nabla^2_{\bmu} \log h(\bmu)
&=-\Sigma^{-1} + \Sigma^{-1} \widetilde{\Sigma} \Sigma^{-1}.
\end{align*}
\end{lemma}
See the proof in the Appendix~\ref{prf: lem gradient orthant prob}.
With the above lemma at hand, we can compute the gradient and Hessian of $-\log\ell(\bbeta_M)$ as
\begin{align*}
-\nabla_{\bbeta_M} \log\ell(\bbeta_M)&=-\Sigma^{-1}(\hat\bbeta_M-\bbeta_M ) + D \Sigma_{\bfb}^{-1}(\widetilde{\bmu_{\bfb}} - \bmu_\bfb )\\
-\nabla^2_{\bbeta_M}\log\ell(\bbeta_M)&=\Sigma^{-1} + D(\Sigma^{-1}_{\bfb}\widetilde{\Sigma_\bfb}\Sigma^{-1}_\bfb-\Sigma^{-1}_\bfb )D,
\end{align*}
where $\widetilde{\bmu_{\bfb}}$ and $\widetilde{\Sigma_{\bfb}}$ are the mean and covariance of the truncated normal distribution {\sloppy$\N(\bmu_\bfb(\bbeta_M),\Sigma_{\bfb})\mid_{\calO}$}.
The SOV method described in Section~\ref{sec: sov} can be used to efficiently evaluate the quantities $\widetilde{\bmu_{\bfb}}$ and $\widetilde{\Sigma_{\bfb}}$. 
Therefore, we can run gradient descent using the SOV estimator of the gradient. After achieving convergence, we compute the Hessian matrix and conduct Wald-type inference the same as described in Section~\ref{sec: intro mle}.

Note that the Hessian matrix $-\nabla^2_{\bbeta_M}\log\ell(\bbeta_M)$ is always positive definite because
\begin{align*}
\Sigma^{-1}\succeq D\Sigma^{-1}_\bfb D,
\end{align*}
since $D\Sigma_{\bfb}D=DH^{-1}D+ \Sigma \succeq \Sigma$. Consequently, the objective function is strongly convex, leading to linear convergence of the gradient descent algorithm. Even though we rely on a gradient estimate rather than the exact gradient in the gradient descent algorithm, this estimate proves to be surprisingly accurate even with a moderate number of RQMC samples. To further enhance the accuracy of the solution, one can consider increasing the number $N$ of RQMC samples as the optimization process approaches the optimum.

\section{Simulations}
\label{sec: simulations}

This section demonstrates the effectiveness of the proposed algorithm via simulations and real data analysis.

\subsection{Carving for linear model}
\label{sec: simu carving}

We adopt a similar experimental setup to the one used in \cite{panigrahi2022approximate}. Specifically, we set $n=300$ and $p=100$. The data are generated as $\bfx_i\iid\N(0,\Sigma_X)$ and $y_i\mid \bfx_i \sim\N(x_i\tran\bbeta,\sigma^2)$ where $\sigma^2=1$. Two types of covariance matrices $\Sigma_X$ are considered: auto-regressive matrix (AR) with $\Sigma_{ij}=0.9^{|i-j|} $ and equi-correlation matrix (Equi) with $\Sigma_{ij}=0.9+0.1\indc{i=j}$. The regression coefficient $\bbeta$ is designed to be a sparse vector with $10$ nonzero coordinates, which are set to be $\pm\sqrt{2c_0\log(p)/n}$ with random signs and with $c_0$ varied among $\{0.6,0.9,1.2\}$. We use 240 observations (i.e. 80\% of the data) for the lasso variable selection, where the lasso regularization parameter $\lambda$ is determined through two different approaches:
\begin{itemize}
\item $\lambda_{\text{CV}}$: choosing $\lambda$ based on a 5-fold cross-validation. Note that the selection of $\lambda$ introduces additional bias but we did not correct for it. This choice is also considered in \cite{schultheiss2021multicarving,panigrahi2022approximate}.

\item $\lambda_{\text{theory}}$: setting $\lambda=\sqrt{\log(p)/n_1}$, as recommended by the theoretical result in \cite{negahban2012unified}. This choice of $\lambda$ is also used by \cite{panigrahi2022approximate}.
\end{itemize}

The objective is to construct 95\% confidence intervals for each coordinate of the target $X_M^\dagger X \bbeta$ in the submodel view. The assessment is based on two key metrics: the average coverage probability, which denotes the proportion of confidence intervals that correctly cover the target parameters, and the average interval lengths. We consider the following inference methods:
\begin{itemize}
\item Splitting: Confidence intervals are constructed using only the remaining 20\% of the data.

\item Bivariate-normal: \cite{panigrahi2022exact} propose to condition even further so that the conditional distribution of $\hat\theta$ becomes tractable. Recall that our inference is based on the conditional distribution given in Equation~\eqref{equ: hat beta cond density}, which marginalizes over the $d$-dimensional vector $\bfb$. To conduct inference for $\beta_{M,j}$, the method in \cite{panigrahi2022exact} conditions on $\bfb_{-j}$ so that the joint conditional distribution of $(\hat\beta_{M,j},b_j)$ is a truncated bivariate normal distribution, hence the name of bivariate-normal. The CDF of this distribution can be precisely evaluated using numerical techniques. However, since the bivariate-normal method conditions on more information than our approach does, we expect it to produce wider confidence intervals.

\item MLE (approx): using the method in \cite{panigrahi2022approximate} to compute the approximate selective MLE and the corresponding Fisher information. We use the implementation available in the GitHub repository\footnote{\url{https://github.com/jonathan-taylor/selective-inference}}.

\item MLE (SOV): employing the SOV-based gradient descent algorithm to find the selective MLE and the corresponding Fisher information, as introduced in Section~\ref{sec: mle}. In each iteration, the gradient is computed using 256 RQMC samples. These RQMC samples are generated from the Sobol' sequence, which is randomized using linear matrix scrambling followed by a digital random shift \citep{matouvsek1998thel2}, using the SciPy package in Python. The step size is set to 0.01 and the algorithm stops when either the change in the log-likelihood or the change in the variable is small enough.

\item CDF (SOV): CDF-based inference using SOV method. A total of 256 RQMC samples are used to construct all the confidence intervals, following the procedure derived in Section~\ref{sec: CI sov} and summarized in Algorithm~\ref{algo}.
\end{itemize}

\begin{figure}
\centering
\begin{subfigure}{.9\textwidth}
\includegraphics[width=\textwidth]{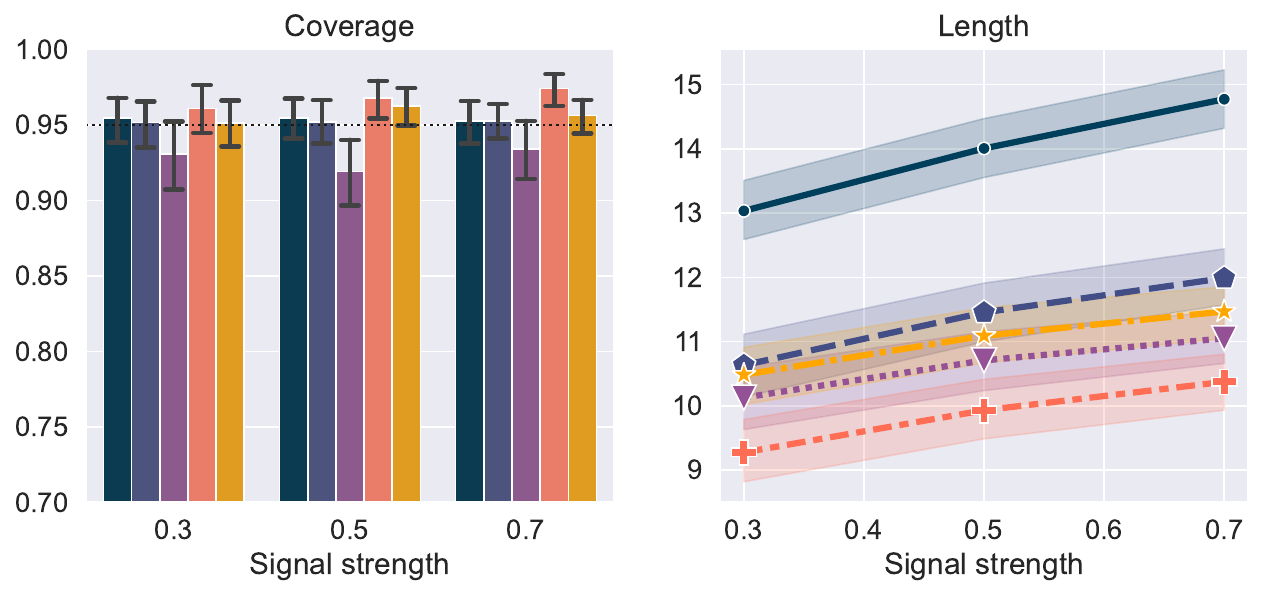}
\caption{Choosing $\lambda=\lambda_{\text{theory}}$}
\end{subfigure}
\begin{subfigure}{.9\textwidth}
\includegraphics[width=\textwidth]{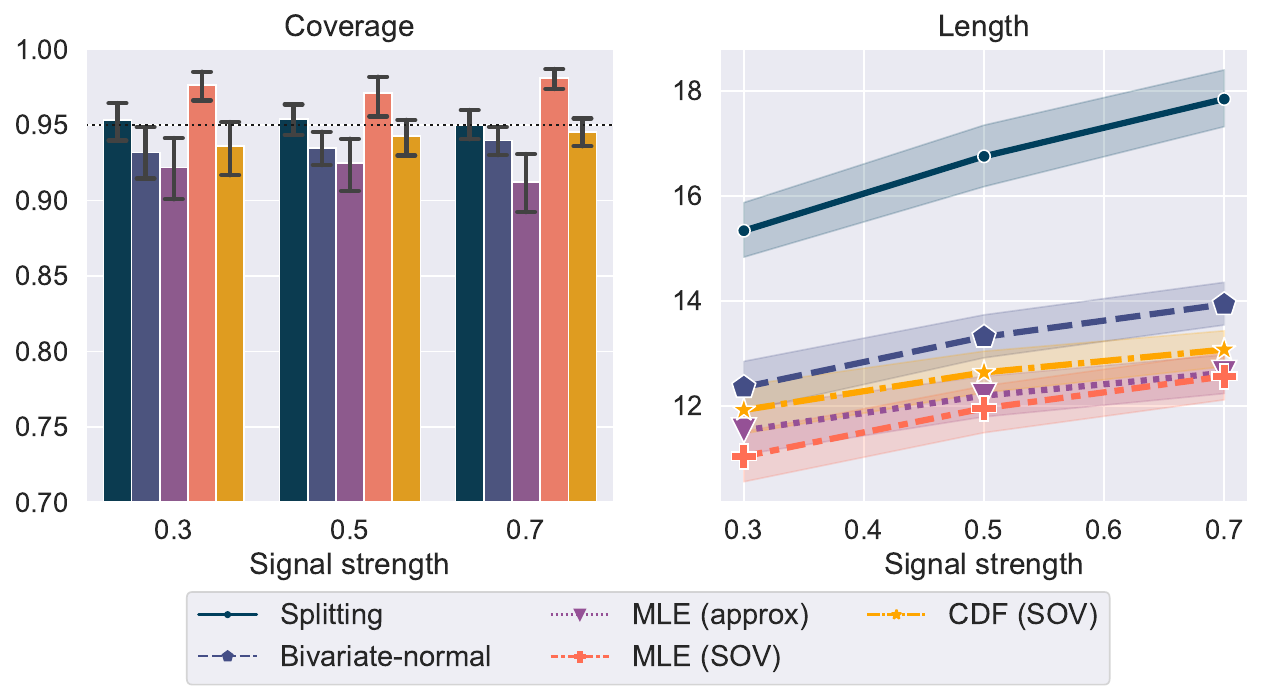}
\caption{Choosing $\lambda=\lambda_{\text{CV}}$.}
\end{subfigure}
\caption{Average coverage probabilities (left panel) and interval lengths (right panel) of intervals constructed by various methods. The targeted coverage probability is 0.95, depicted by the dotted line. The $x$-axes represent the signal strength $c_0$. In the left panel, the 5 methods from left to right are Splitting, Bivariate-normal, MLE (approx), MLE (SOV), and CDF (SOV). The error bars represent 95\% confidence intervals produced by bootstrapping from the 200 repeated simulations. The regularization parameter $\lambda$ is selected based on the theory (top panel) or through cross-validation (bottom panel). The covariance matrix $\Sigma_X$ of the features $\bfx_i$ is the auto-regressive matrix with $(i,j)$ entry being $0.9^{|i-j|}$. 
}
\label{fig: AR1}
\end{figure}

\begin{figure}
\centering
\begin{subfigure}{.9\textwidth}
\includegraphics[width=\textwidth]{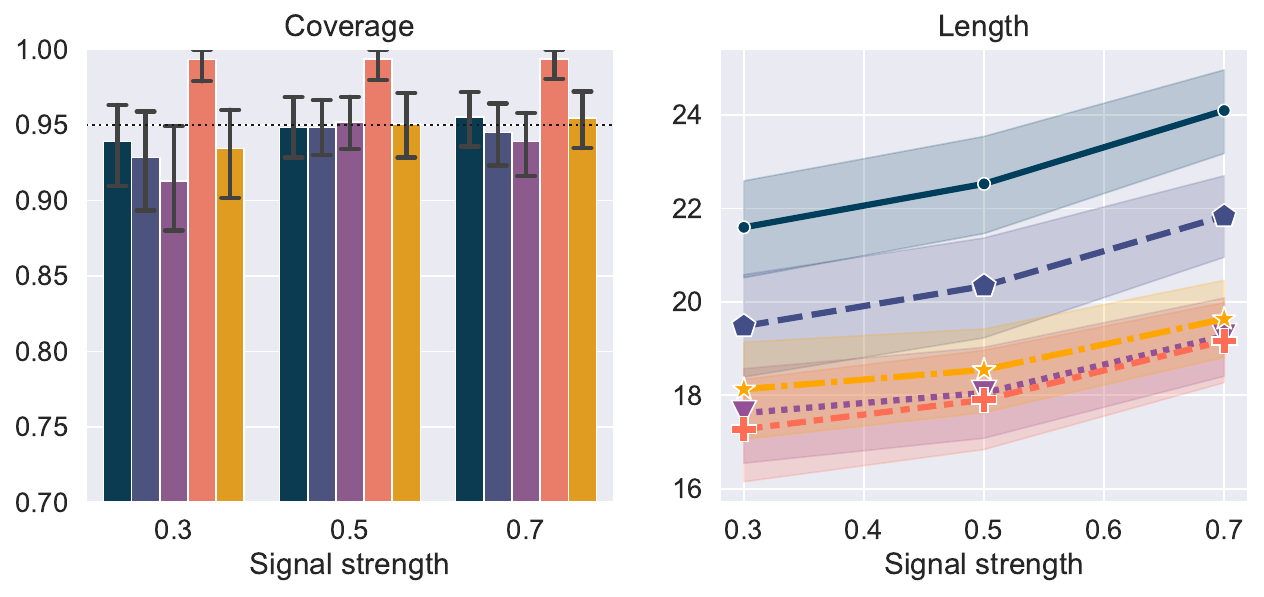}
\caption{Choosing $\lambda=\lambda_{\text{theory}}$.}
\end{subfigure}
\begin{subfigure}{.9\textwidth}
\includegraphics[width=\textwidth]{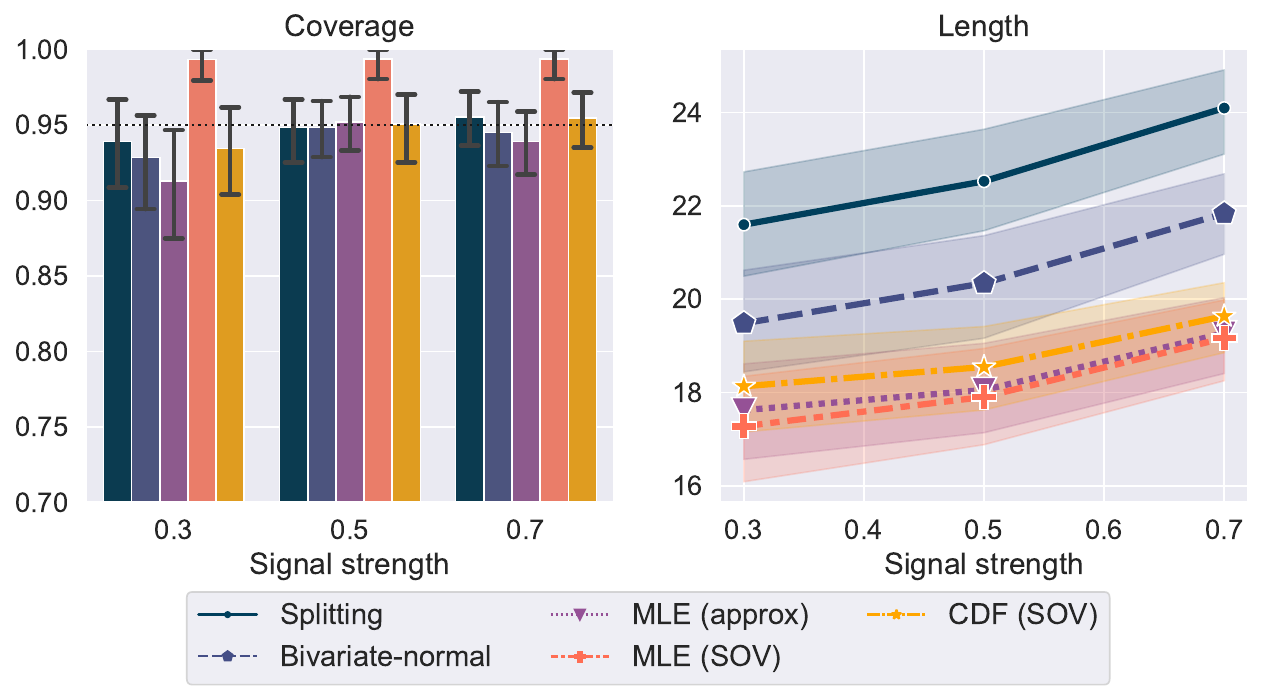}
\caption{Choosing $\lambda=\lambda_{\text{CV}}$.}
\end{subfigure}
\caption{The protocol is the same as Figure~\ref{fig: AR1}, except that $\Sigma_X$ is the equi-correlation matrix with $(i,j)$ entry being $0.9+0.1\indc{i=j}$.}
\label{fig: equi}
\end{figure}

For all these methods, $\sigma^2$ is estimated by 
\begin{align*}
    \hat\sigma^2=\frac{1}{n-p}\|Y-X(X\tran X)^{-1}X\tran Y\|_2^2.
\end{align*}
The simulation is repeated 200 times and the results are presented in Figures~\ref{fig: AR1} and \ref{fig: equi}. In Figure~\ref{fig: AR1}, the covariance $\Sigma_X$ is the auto-regressive (AR) matrix with $(i,j)$ entry being $0.9^{|i-j|}$, while in Figure~\ref{fig: equi} it is the equi-correlation (Equi) matrix with $(i,j)$ entry being $0.9+0.1\indc{i=j}$. In both Figures, the top panels choose $\lambda=\lambda_{\text{theory}}$ while the bottom panels choose $\lambda=\lambda_{\text{CV}}$. The $x$-axes represent the signal strength $c_0$. The error bars are the 95\% confidence intervals produced by bootstrapping from the 200 repetitions.
A few observations are in order:
\begin{enumerate}[(1)]
\item The coverage probabilities of MLE (approx) method tend to fall short of the desired 0.95 coverage, especially when $\lambda$ is selected by cross-validation. This discrepancy can be attributed to the fact that the MLE (approx) method only provides an approximate estimation of the selective MLE. In situations where the approximation is not accurate, the reliability of the method diminishes. On the contrary, the proposed MLE (SOV) method consistently achieves or even surpasses the targeted coverage probability.  

\item The CDF-based methods, namely Splitting, Bivariate-normal, and CDF (SOV), attain the desired coverage probabilities across all scenarios.

\item In terms of interval lengths, we observe a consistent order among the methods: Splitting $>$ Bivariate-normal $>$ SOV+IS $>$ MLE (approx) $>$ MLE (SOV) across all scenarios. Splitting uses the least amount of information for inference, consequently producing the longest intervals. The Bivariate-normal method, which also conditions on more information than the remaining methods, yields longer intervals than others, especially in the strong correlation scenario as shown in Figure~\ref{fig: equi}. This is because when stronger correlations exist among variables, conditioning on $\bfb_{-j}$ leaves less information in $b_j$, resulting in longer intervals. The MLE-based methods do not condition on the nuisance parameters as the CDF-based methods do, thus the two MLE-based methods are the shortest as expected. Interestingly, the MLE (SOV) intervals are slightly shorter than the MLE (approx) intervals, despite MLE (SOV) having higher coverage probabilities.

\end{enumerate}

The wall clock times of these methods are presented in the Appendix~\ref{appendix: computation}. While the sampling-based method entails a slightly higher computational cost compared to the approximate MLE method, which only solves a convex problem, the average inference time for each experiment is less than one second. This indicates that our method offers more reliable and accurate inference with only a minimal increase in computation cost.

\subsection{Compare hit-and-run and SOV}

To demonstrate the superiority of the proposed SOV method to the previously used hit-and-run algorithm \citep{belisle1993hit}, we will compare their coverage probabilities, interval lengths, as well as wall clock times. The setting is the same as that in the bottom panel of Figure~\ref{fig: AR1}, where $\lambda$ is selected by cross-validation and $\Sigma_X$ is the auto-regressive matrix. The signal strength is fixed to $c_0=0.7$. The $x$-axis represents the number of samples $N$ used by the two sampling methods. Note that the hit-and-run sampler uses an extra 20 samples as burn-in.

The hit-and-run sampler is used to sample $\bfb$ from $\N(\bmu_{\bfb}(\bfeta), \Sigma_\bfb(\bfeta) )\mid_{\calO}$ given in Corollary~\ref{coro: carving Sigma_omega}, with $\theta$ set to be $\hat\theta^{(2)}$. Here, $\hat\theta^{(2)}$ is the unbiased MLE of $\theta$ obtained using the hold-out 20\% data. We apply importance weighting to evaluate the CDF at different values of $\theta$ and to compute the confidence intervals. The hit-and-run sampler is initialized at the mode of the target truncated Gaussian distribution. Finding the mode is a straightforward convex problem. The sampler moves not only in the coordinate direction, but also in the leading principal component (PC) direction of the covariance matrix of the Gaussian distribution. We use the first $k$ PCs such that these components explain over half of the variance. Moving along the PCs allow the sampler to explore the entire distribution more efficiently, especially when there exist strong correlations. The computational overhead of finding the mode and the PCs are not factored into the computation time of hit-and-run.

\begin{figure}
\centering
\includegraphics[width=.9\textwidth]{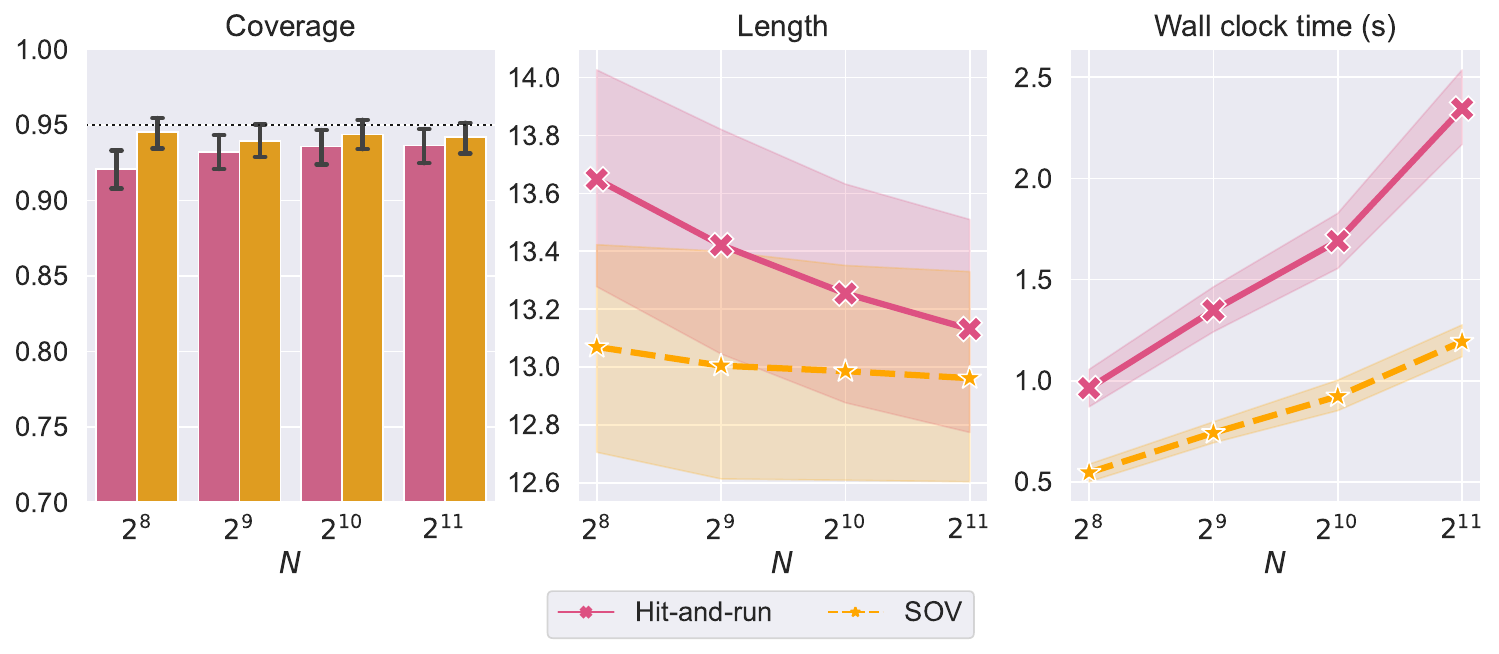}
\caption{Compare the proposed SOV method with hit-and-run in terms of coverage probabilities (left panel), interval lengths (middle panel), and wall clock time (right panel). The experimental setting is the same as the bottom panel of Figure~\ref{fig: AR1}. }
\label{fig: hnr vs sov}
\end{figure}

The results are shown in Figure~\ref{fig: hnr vs sov}. A few observations are in order:
\begin{enumerate}[(1)]
\item When the sample size $N$ is small, the hit-and-run method suffers from under-coverage, while the SOV method achieves the desired coverage even with 256 samples. As $N$ increase, the hit-and-run method eventually achieves the same coverage as the SOV method.

\item For small $N$, the intervals generated by the hit-and-run method exhibit not only lower coverage probabilities but also longer lengths. This phenomenon could potentially be attributed to the fact that the hit-and-run algorithm is initialized at the mode of $\N(\bmu_{\bfb}(\bfeta), \Sigma_\bfb(\bfeta) )\mid_{\calO}$, and $\Phi(\frac{\hat\theta - \mu_{\hat\theta}(\bfb)}{\sigma_{\hat\theta}} )$ tends to be larger when $\bfb$ is at the mode than on average. Consequently, when the samples are concentrated around the mode, the resulting p-values tend to be biased upwards, subsequently leading to longer confidence intervals.  As $N$ increases, the hit-and-run sampler explores the entire distribution more sufficiently and the bias diminishes, yielding more accurate confidence intervals.

\item The right panel shows that the SOV method is much faster compared to hit-and-run when comparing wall clock times.  This highlights the greater efficiency of the SOV method in comparison to hit-and-run.
\end{enumerate}

\subsection{Real data analysis}

We consider the HIV drug resistance data from \cite{rhee2003human}, which explores the predictive potential of various mutations for drug resistance in HIV. The response is the log susceptibility to the drug and the predictors consist of the mutations. Following \cite{panigrahi2021integrative}, we focus on the drug 3TC and discard the mutations occurring fewer than 10 times in the data. This leads to a dataset of size $n=633$ and $p=91$. We apply the lasso on a random subset of 80\% of the data, resulting in 17 selected mutations. 

The next goal is to compute p-values for testing whether each of selected parameter is null. These p-values are computed using the method outlined in Section~\ref{sec: cdf inference}. Our primary focus is to compare the proposed SOV method versus the hit-and-run method. The SOV method uses $2^{12}$ RQMC samples, while the hit-and-run method generates $5\times 2^{12}$ samples, with an additional 20 samples designated for burn-in. Since both Monte Carlo estimators are random, we repeat the computation 50 times with independent randomization and use the sample standard error as an estimate of the error of the estimator.

Table~\ref{table: hiv} presents the computed p-values, accompanied by their respective error estimates enclosed in parentheses. Four variables are ignored because their p-values are extremely close to zero (smaller than $10^{-6}$) indicated by both methods. The p-values obtained by the two methods are close to each other. However, the errors of the SOV estimator are several orders of magnitude smaller than those of the hit-and-run method, even though the latter employs 5 times more samples. This highlights the remarkable precision achieved by the SOV method in contrast to the hit-and-run method.
\begin{table}
\centering
\begin{tabular}{c|cc}
    \toprule
Mutations &          Hit-and-run &                  SOV  \\
    \midrule
    P41L  &  0.98 (0.0149) &  0.99 (0.0003) \\
    P62V  &  0.26 (0.0165) &  0.26 (0.0007) \\
    P75I  &  0.70 (0.0261) &  0.69 (0.0004) \\
    P75T  &  0.66 (0.0136) &  0.66 (0.0002) \\
    P77L  &  0.47 (0.0275) &  0.47 (0.0002) \\
    P83K  &  0.02 (0.0046) &  0.02 (0.0001) \\
    P115F &  0.12 (0.0134) &  0.12 (0.0001) \\
    P118I &  0.36 (0.0111) &  0.36 (0.0001) \\
    P151M &  0.12 (0.0182) &  0.12 (0.0002) \\
    P219R &  0.06 (0.0058) &  0.06 (0.0001) \\
    P210W &  0.24 (0.0110) &  0.24 (0.0001) \\
    P215Y &  0.006346 (0.00167) &  0.006428 (0.00003) \\
    P181C &  0.000073 (0.00009) &  0.000097 (0.00002) \\
    \bottomrule
    \end{tabular}
\caption{P-values for the selected variables in the HIV data. Errors are shown in parentheses and were estimated as the standard error among 50 random replicates.}
\label{table: hiv}
\end{table}

\section{Conclusion}
\label{sec: conclusion}

Conducting conditional selective inference is often challenging due to the complexity of the conditional distributions involved. This paper developed an efficient method for sampling from such distributions, in scenarios where the selection event can be characterized by a polyhedron. Moreover, the method can be employed to compute the maximum of the selection-adjusted likelihood, facilitating efficient MLE-based inference. Empirical evaluations were performed, comparing the method against various recently proposed approaches, in the context of the randomized lasso problem.

Although primarily illustrated within the lasso framework, the methodology can be applied to other scenarios involving polyhedral selection as highlighted in the introduction. In cases with unknown covariance, the conditional distribution may involve the orthant probability of a multivariate t-distribution, for which a similar SOV method can be employed \citep{genz1999numerical}.

\clearpage
\appendix
\section{Proofs}

\subsection{Proof of Proposition~\ref{prop: cdf}}
\label{proof: prop cdf}
\begin{proof}
The joint unconditional density of $(\hat\theta,\bomega)$ is the product of Gaussian
\begin{align*}
p(\hat\theta, \omega)= \varphi(\hat\theta;\theta, \nu )\cdot \varphi(\bomega; 0, \Omega ).
\end{align*}
Applying the change-of-variable formula from $(\hat\theta,\omega)$ to $(\hat\theta,\bfb,\bfs_{-M})$ while conditioning on $\{\calE,\bbeta^\perp\}$, we get the joint density
\begin{align*}
p(\hat\theta,\bfb\mid\calE,\bbeta^\perp)&\propto \Exp{-\frac1{2\nu}(\hat\theta-\theta)^2 - \frac12(\wtQ_1\hat\theta+Q_2 \bfb +\bft)\tran \Omega^{-1}(\wtQ_1\hat\theta+Q_2 \bfb+\bft) }\\
&\propto \Exp{-\frac12 (\nu^{-1}+\wtQ_1\tran\Omega^{-1}\wtQ_1) \hat\theta^2 - \frac12 \bfb\tran Q_2\tran \Omega^{-1} Q_2\bfb + \hat\theta (\nu^{-1}\theta - \wtQ_1\tran\Omega^{-1} \bft) - \bfb\tran Q_2\tran\Omega^{-1}\bft - \hat\theta \wtQ_1\tran \Omega^{-1} Q_2\bfb }.
\end{align*}
Let
\begin{align*}
\sigma_{\hat\theta}^2&=\frac{1}{\nu^{-1} + \wtQ_1\tran\Omega^{-1}\wtQ_1},\\
\mu_{\hat\theta}&=\sigma_{\hat\theta}^2 (\nu^{-1}\theta - \wtQ_1\tran\Sigma_{\Omega}^{-1}\bft - \wtQ_1\tran\Omega^{-1}Q_2 \bfb ).
\end{align*} 
Then the above density is proportional to
\begin{align*}
\varphi(\hat\theta; \mu_{\hat\theta}, \sigma_{\hat\theta}^2 ) \cdot \Exp{-\frac12\bfb\tran(Q_2\tran\Omega^{-1}Q_2 -\sigma_{\hat\theta}^2 Q_2\tran\Omega^{-1}\wtQ_1\wtQ_1\tran\Omega^{-1}Q_2 )\bfb + \bfb\tran(-Q_2\tran\Omega^{-1}\bft - \sigma_{\hat\theta}^2 Q_2\tran\Omega^{-1}\wtQ_1(\nu^{-1}\theta - \wtQ_1\tran\Omega^{-1}\bft ) ) }
\end{align*}
Denote 
\begin{align*}
\Sigma_{\bfb}&=(Q_2\tran\Omega^{-1}Q_2 -\sigma_{\hat\theta}^2 Q_2\tran\Omega^{-1}\wtQ_1\wtQ_1\tran\Omega^{-1}Q_2 )^{-1},\\
\bmu_\bfb&=\Sigma_{\bfb}(-Q_2\tran\Omega^{-1}\bft - \sigma_{\hat\theta}^2 Q_2\tran\Omega^{-1}\wtQ_1(\nu^{-1}\theta - \wtQ_1\tran\Omega^{-1}\bft ) ).
\end{align*}
Therefore, the conditional density of $(\hat\theta,\bfb)\mid\{\calE,\bbeta^\perp\}$ is proportional to
\begin{align*}
p(\hat\theta, \bfb\mid \calC)&\propto  \varphi(\hat\theta; \mu_{\hat\theta},\sigma^2_{\hat\theta} )\cdot \varphi(\bfb;\bmu_{\bfb}, \Sigma_\bfb )\cdot \Indc{\bfb\in\calO }.
\end{align*}

Note that $\wtQ_1=-Q_2\tilde{\bfc}$. So $\wtQ_1\tran\Omega^{-1}\wtQ_1=\tilde{\bfc}\tran H\tilde{\bfc}$, $\wtQ_1\tran \Omega^{-1}=-\tilde{\bfc}\tran Q_2\tran \Omega^{-1} $, $\wtQ_1\tran\Omega^{-1}Q_2=-\tilde{\bfc}\tran H$. 
By the Sherman-Morrison formula, $\Sigma_{\bfb}$ can be simplified as
\begin{align*}
\Sigma_{\bfb}&=(H-\sigma_{\hat\theta}^2 H\tilde{\bfc} \tilde{\bfc}\tran H)^{-1}=H^{-1} + \frac{\sigma_{\hat\theta}^2}{1-\sigma_{\hat\theta}^2 \tilde{\bfc}\tran H \tilde{\bfc} } \tilde{\bfc} \tilde{\bfc}\tran\\
&=H^{-1}+\frac{1}{\sigma_{\hat\theta}^{-2} - \tilde{\bfc}\tran H\tilde{\bfc}  }\tilde{\bfc} \tilde{\bfc}\tran\\
&=H^{-1}+\nu \tilde{\bfc} \tilde{\bfc}\tran.
\end{align*}

Moreover,
\begin{align*}
\Sigma_{\bfb}^{-1}\bmu_{\bfb}&=-Q_2\tran\Omega^{-1}\bft - \sigma_{\hat\theta}^2 Q_2\tran\Omega^{-1}\wtQ_1(\nu^{-1}\theta - \wtQ_1\tran\Omega^{-1}\bft )\\
&=-\bfk+\sigma_{\hat\theta}^2 H\tilde{\bfc}(\nu^{-1} + \tilde{\bfc}\tran Q_2\tran\Omega^{-1}\bft )\\
&=-\bfk+\sigma_{\hat\theta}^2 H\tilde{\bfc}(\nu^{-1} + \tilde{\bfc}\tran\bfk ).
\end{align*}
\end{proof}

\subsection{Proof of Proposition~\ref{prop: selective likelihood}}
\label{prf: selective likelihood}
\begin{proof}
It suffices to show that the denominator in \eqref{equ: hat beta cond density} is proportional to (up to constants that do not depend on $\bbeta_M$)
\begin{align*}
\int_{\bfb\in\calO}\varphi(\bfb;\mu_{\bfb},\Sigma_{\bfb} )\rd \bfb.
\end{align*}
To show this, note that
\begin{align*}
&\varphi(\hat\bbeta_M;\bbeta_M, \Sigma )\cdot \varphi(Q_1\hat\bbeta_M + Q_2\bfb + \bfr+\bfs;\bzero,\Omega )\\
&\quad \propto \Exp{-\frac12 (\hat\bbeta_M - \bbeta_M)\tran\Sigma^{-1}(\hat\bbeta_M - \bbeta_M) - \frac12 (Q_1\hat\bbeta_M + Q_2\bfb + \bfr+\bfs )\tran \Omega^{-1} (Q_1\hat\bbeta_M + Q_2\bfb + \bfr+\bfs) }\\
&\quad =\mathrm{exp}\left\{-\frac12\hat\bbeta_M\tran (\Sigma^{-1} + Q_1\tran\Omega^{-1}Q_1 ) \hat\bbeta_M + \hat\bbeta_M\tran(\Sigma^{-1}\bbeta_M - Q_1\tran \Omega^{-1}(Q_2\bfb+\bfr+\bfs) )  \right.\\
&\qquad \left. -\frac12 \bfb\tran Q_2\tran\Omega^{-1}Q_2 \bfb - \bfb\tran Q_2\tran\Omega^{-1}(\bfr+\bfs) - \frac12 \bbeta_M\tran \Sigma^{-1}\bbeta_M  \right\}.
\end{align*}
Let
\begin{align*}
\Lambda&=(\Sigma^{-1}+Q_1\tran\Omega^{-1}Q_1)^{-1}=(\Sigma^{-1}+DHD)^{-1} ,\\
\bfm&=\Lambda(\Sigma^{-1}\bbeta_M - Q_1\tran \Omega^{-1}(Q_2\bfb+\bfr+\bfs) ).
\end{align*}
Then the above display is proportional to
\begin{align*}
\varphi(\hat\bbeta_M; \bfm,\Lambda  ) \cdot &\Exp{\frac12(\Sigma^{-1}\bbeta_M - Q_1\tran \Omega^{-1}(Q_2\bfb+\bfr+\bfs) )\tran \Lambda(\Sigma^{-1}\bbeta_M - Q_1\tran \Omega^{-1}(Q_2\bfb+\bfr+\bfs) )
\right.\\
&\qquad\left. -\frac12 \bfb\tran Q_2\tran\Omega^{-1}Q_2 \bfb - \bfb\tran Q_2\tran\Omega^{-1}(\bfr+\bfs) -\frac12\bbeta_M\tran \Sigma^{-1}\bbeta_M  }.
\end{align*}
Because only the first term $\varphi(\hat\bbeta_M; \bfm,\Lambda  )$ depends on $\hat\bbeta_M$ and it is a density, so it vanishes when we integrate $\hat\bbeta_M$ over $\bbR^d$. The remaining term becomes
\begin{align*}
&\Exp{-\frac12 \bfb\tran (Q_2\tran\Omega^{-1}Q_2 - Q_2\tran\Omega^{-1}Q_1\Lambda Q_1\tran\Omega^{-1}Q_2 )\bfb  \right.\\
&\qquad \left. +\bfb\tran(-Q_2\tran\Omega^{-1}Q_1\Lambda(\Sigma^{-1}\bbeta_M - Q_1\tran\Omega^{-1}(\bfr+\bfs) ) - Q_2\tran\Omega^{-1}(\bfr+\bfs)) \right.\\
&\qquad \left. + \frac12 \bbeta_M\tran \Sigma^{-1}\Lambda \Sigma^{-1} \bbeta_M -\frac12 \bbeta_M\tran \Sigma^{-1}\bbeta_M -\bbeta_M\tran \Sigma^{-1}\Lambda Q_1\tran\Omega^{-1} (\bfr+\bfs ) 
}.\numberthis\label{equ: prf mle}
\end{align*}
Let
\begin{align*}
\Sigma_\bfb &= (Q_2\tran\Omega^{-1}Q_2 - Q_2\tran\Omega^{-1}Q_1\Lambda Q_1\tran\Omega^{-1}Q_2 )^{-1}, \\
\mu_{\bfb}&=\Sigma_\bfb(-Q_2\tran\Omega^{-1}Q_1\Lambda(\Sigma^{-1}\bbeta_M - Q_1\tran\Omega^{-1}(\bfr+\bfs) ) - Q_2\tran\Omega^{-1}(\bfr+\bfs) ).
\end{align*}

Recall that $H=Q_2\tran\Omega^{-1}Q_2$, $Q_1=-Q_2D$, $Q_1\tran\Omega^{-1}Q_1=DHD$.
By the Woodbury matrix identity,
\begin{align*}
\Sigma_\bfb=(H-HD\Lambda DH )^{-1}=H^{-1}+D (\Lambda^{-1} - DHD )^{-1}D=H^{-1}+D\Sigma D.
\end{align*}
Moreover,
\begin{align*}
\Sigma_\bfb^{-1}\mu_\bfb&=HD\Lambda\Sigma^{-1}\bbeta_M +(Q_2\tran\Omega^{-1}Q_1\Lambda Q_1\tran\Omega^{-1}- Q_2\tran\Omega^{-1}) (\bfr+\bfs)\\
&=HD\Lambda\Sigma^{-1}\bbeta_M +(HD\Lambda D - I) Q_2\tran\Omega^{-1}(\bfr+\bfs).
\end{align*}
Since $\Sigma_\bfb^{-1}H^{-1}=I-HD\Lambda D$,
\begin{align*}
\bmu_\bfb&=\Sigma_\bfb HD\Lambda\Sigma^{-1}\bbeta_M -H^{-1}Q_2\tran \Omega^{-1}(\bfr+\bfs).
\end{align*}
Also note that
\begin{align*}
\Sigma_\bfb HD\Lambda\Sigma^{-1}=(I+D\Sigma DH) D (\Sigma^{-1}+DHD)^{-1}\Sigma^{-1}=D(I+\Sigma DHD) (I+\Sigma DHD)^{-1}=D.
\end{align*}
Thus
\begin{align*}
\bmu_{\bfb}=D \bbeta_M - H^{-1}Q_2\tran\Omega^{-1}(\bfr+\bfs).
\end{align*}
We also have
\begin{align*}
D\Sigma_\bfb^{-1}D=DHD - DHD\Lambda DHD=(\Lambda^{-1}-\Sigma^{-1})-(\Lambda^{-1}-\Sigma^{-1}) \Lambda (\Lambda^{-1}-\Sigma^{-1}) = \Sigma^{-1}-\Sigma^{-1}\Lambda \Sigma^{-1},
\end{align*}
which shows that $\Sigma^{-1}\succeq D\Sigma_\bfb^{-1}D$.

The density in Equation~\eqref{equ: prf mle} is proportional to
\begin{align}
\label{equ: prf mle 2}
&\varphi(\bfb;\bmu_\bfb,\Sigma_\bfb )\cdot \Exp{\frac12 \bmu_\bfb\tran \Sigma_\bfb^{-1}\bmu_{\bfb}  
+ \frac12 \bbeta_M\tran (\Sigma^{-1}\Lambda \Sigma^{-1}-\Sigma^{-1}) \bbeta_M + \bbeta_M\tran\Sigma^{-1}\Lambda D Q_2\tran\Omega^{-1}(\bfr+\bfs) }.
\end{align}
Note that
\begin{align*}
\frac12\bmu_\bfb\Sigma_\bfb^{-1}\bmu_\bfb&=\frac12 (D\bbeta_M - H^{-1}Q_2\tran\Omega^{-1}(\bfr+\bfs))\tran \Sigma_\bfb^{-1}(D\bbeta_M - H^{-1}Q_2\tran\Omega^{-1}(\bfr+\bfs))\\
&=\frac12 \bbeta_M\tran D\Sigma_\bfb^{-1}D \bbeta_M - \bbeta_M\tran D\Sigma_\bfb^{-1} H^{-1}Q_2\tran\Omega^{-1}(\bfr+\bfs) + \text{constant}\\
&=\frac12 \bbeta_M\tran(\Sigma^{-1}-\Sigma^{-1}\Lambda \Sigma^{-1} )\bbeta_M - \bbeta_M\tran D \Sigma_\bfb^{-1}H^{-1}Q_2\tran\Omega^{-1} (\bfr+\bfs)+\text{constant}\\
&=\frac12 \bbeta_M\tran(\Sigma^{-1}-\Sigma^{-1}\Lambda \Sigma^{-1} )\bbeta_M - \bbeta_M\tran \Sigma^{-1}\Lambda D Q_2\tran\Omega^{-1} (\bfr+\bfs)+\text{constant}
\end{align*}
This shows that the quantity
\begin{align*}
\Exp{\frac12 \bmu_\bfb\tran \Sigma_\bfb^{-1}\bmu_{\bfb}  
+ \frac12 \bbeta_M\tran (\Sigma^{-1}\Lambda \Sigma^{-1}-\Sigma^{-1}) \bbeta_M + \bbeta_M\tran\Sigma^{-1}\Lambda D Q_2\tran\Omega^{-1}(\bfr+\bfs) }
\end{align*}
is constant in $\bbeta_M$.
Therefore, the quantity in Equation~\eqref{equ: prf mle 2} is proportional to $\varphi(\bfb;\bmu_\bfb,\Sigma_\bfb)$.
That is, the denominator in \eqref{equ: hat beta cond density} is proportional to
\[
\int_{\bfb\in\calO}\varphi(\bfb;\mu_\bfb,\Sigma_\bfb)\rd \bfb.
\]

In the special case when $\Omega=\kappa^{-1}\cdot \sigma^2 X\tran X$,
we have
$Q_2\tran\Omega^{-1}= DX_M\tran X (\kappa/\sigma^2) (X\tran X)^{-1} = (\kappa/\sigma^2)D J_M $, in which case
\begin{align*}
\bmu_\bfb =D\bbeta_M - (\kappa/\sigma^2) H^{-1} D (\bfr_M+\bfs_M). 
\end{align*}



\end{proof}

\subsection{Proof of Lemma~\ref{lem: importance weight}}
\label{prf: lem importance weight}
\begin{proof}
The importance weight is proportional to
\begin{align*}
&\Exp{-\frac12 \bfb\tran (\Sigma_{\bfb}(\bfeta)^{-1} - H ) \bfb + \bfb\tran (\Sigma_\bfb(\bfeta)^{-1}\bmu_{\bfb}(\bfeta,\theta) - \bar\Sigma^{-1}\bar\bmu  )    }\\
&\qquad=\Exp{\frac{\sigma_{\hat\theta}^2}{2}  (\bfb\tran H\tilde{\bfc})^2 + \bfb\tran (\Sigma_\bfb(\bfeta)^{-1}\bmu_{\bfb}(\bfeta,\theta) - \bar\Sigma^{-1}\bar\bmu  )  }.
\end{align*}
Note that
\begin{align*}
\Sigma_{\bfb}(\bfeta)^{-1}\bmu_{\bfb}(\bfeta,\theta) - \bar\Sigma^{-1}\bar\bmu &=-\bfk + \sigma^2_{\hat\theta}H\tilde{\bfc}(\nu^{-1}\theta + \tilde{\bfc}\tran\bfk) + Q_2\tran\Omega^{-1}(\bfr+\bfs - X\tran X_M\hat\bbeta_M )\\
&=\sigma^2_{\hat\theta}H\tilde{\bfc}(\nu^{-1}\theta + \tilde{\bfc}\tran\bfk) - Q_2\tran\Omega^{-1}X\tran X_M \bfc \hat\theta\\
&=\sigma^2_{\hat\theta}H\tilde{\bfc}(\nu^{-1}\theta + \tilde{\bfc}\tran\bfk) - H\tilde{\bfc}\hat\theta\\
&=\sigma^2_{\hat\theta} H\tilde{\bfc}(\nu^{-1}\theta + \tilde{\bfc}\tran\bfk - \frac{\hat\theta}{\sigma_{\hat\theta}^2} )\\
&=\Delta \btau.
\end{align*}
\end{proof}

\subsection{Proof of Lemma~\ref{lemma: gradient orthant prob}}
\label{prf: lem gradient orthant prob}
\begin{proof}
Note that 
\begin{align*}
\nabla_{\bmu}\log h(\bmu)&=\frac{1}{h(\bmu)} \nabla_{\bmu} h(\bmu)=\frac{1}{h(\bmu)}\nabla_{\bmu}\int_{\bfb\in\calO} \varphi(\bfb;\bmu,\Sigma )\rd\bfb\\
&=\frac{1}{h(\bmu)} \int_{\bfb\in\calO} \Sigma^{-1}(\bfb-\bmu )  \varphi(\bfb;\bmu,\Sigma )\rd\bfb\\
&=\Sigma^{-1}\frac{ \int_{\bfb\in\calO} (\bfb-\bmu )  \varphi(\bfb;\bmu,\Sigma )\rd\bfb}{\int_{\bfb\in\calO}   \varphi(\bfb;\bmu,\Sigma )\rd\bfb}\\
&=\Sigma^{-1}(\tilde{\bmu} - \bmu ).
\end{align*}
Similarly,
\begin{align*}
\nabla^2_{\bmu} \log h(\bmu)
&=\frac{1}{h(\bmu)}\left[\Sigma^{-1}\int_{\calO}(\bfb-\bmu)(\bfb-\bmu)\tran \varphi(\bfb;\bmu,\Sigma)\rd \bfb\Sigma^{-1} - \Sigma^{-1} h(\bmu) \right] - \nabla\log h(\bmu) \nabla \log h(\bmu)\tran\\
&=\Sigma^{-1} [\widetilde{\Sigma} + (\tilde{\bmu} - \bmu )(\tilde{\bmu} - \bmu)\tran ]\Sigma^{-1} -\Sigma^{-1} - \Sigma^{-1}(\tilde{\bmu} - \bmu )(\tilde{\bmu} - \bmu)\tran \Sigma^{-1} \\
&=-\Sigma^{-1} + \Sigma^{-1} \widetilde{\Sigma} \Sigma^{-1}.
\end{align*}
\end{proof}

\section{Computation cost}
\label{appendix: computation}
Table~\ref{table: time} shows the computation time in seconds of various methods in various settings. The MLE (SOV) and CDF (SOV) methods are more time-consuming. However, the computation is still reasonably fast. More importantly, the MLE (SOV) and CDF (SOV) methods are more reliable compared to MLE (approx) as shown in Section~\ref{sec: simu carving}.
\begin{table}
\centering
\begin{tabular}{cc|ccc}
\toprule
$\lambda$ & $\Sigma_X$ &   MLE (approx) &  MLE (SOV) &   CDF (SOV) \\
\midrule
$\lambda_{\text{CV}}$ & AR  &      0.0496 &   0.5788 & 0.4356 \\
$\lambda_{\text{CV}}$ & Equi  &      0.0582 &   0.1664 & 0.2605 \\
$\lambda_{\text{theory}}$ & AR  &      0.0470 &   0.3789 & 0.2248 \\
$\lambda_{\text{theory}}$ & Equi   &      0.0515 &   0.1382 & 0.1652 \\
\bottomrule
\end{tabular}
\caption{Wall clock time in seconds of different methods for different choices of $\lambda$ and the covariance $\Sigma_X$. The experiment settings are the same as in Section~\ref{sec: simu carving}. All computations were conducted on a computer node equipped with 2 CPUs, each with 4GB of memory.}
\label{table: time}
\end{table}

\bibliography{arxiv.bbl}
\bibliographystyle{apalike}

\end{document}